\documentclass[useAMS,usenatbib]{mn2e}

\usepackage{graphicx}
\usepackage{verbatim}
\usepackage{fix2col}

\title[Multi-wavelength properties of SDSS galaxy classes]{The nature of the SDSS galaxies in various classes based on morphology, colour and spectral features -- II. Multi-wavelength properties}
\author[J. H. Lee et al.]{Joon Hyeop Lee$^{1,2}$\thanks{E-mail:
jhl@kasi.re.kr (JHL); mglee@astro.snu.ac.kr (MGL); cbp@kias.re.kr (CBP); yychoi@kias.re.kr (YYC)}, Myung Gyoon Lee$^{2\star}$, Changbom Park$^{3\star}$, Yun-Young Choi$^{4\star}$\\
$^{1}$Korea Astronomy and Space Science Institute, Daejeon 305-348, Korea\\
$^{2}$Astronomy Program, Department of Physics and Astronomy, Seoul National University, Seoul 151-742, Korea\\
$^{3}$Korea Institute for Advanced Study, Dongdaemun-gu, Seoul 103-722, Korea\\
$^{4}$Astrophysical Research Centre for the Structure and Evolution of the Cosmos, Sejong University, Seoul 143-747, Korea}
\begin{document}

\date{Accepted 2009 September 21. Received 2009 September 21; in original form 2008 August 30}

\pagerange{\pageref{firstpage}--\pageref{lastpage}} \pubyear{2008}

\maketitle

\label{firstpage}

\begin{abstract}
We present a multi-wavelength study of the nature of the SDSS galaxies 
divided into fine classes based on their morphology, colour and spectral features.
The SDSS galaxies are classified into early-type and late-type; red and blue;
passive, H{\protect\scriptsize II}, Seyfert and LINER, which returns a total of 16 fine classes of galaxies.
The properties of galaxies in each fine class are investigated from radio to X-ray, using 2MASS, IRAS, FIRST, NVSS, GALEX and ROSAT data.
The UV -- optical -- NIR colours of blue early-type galaxies (BEGs) seem to result from the combination of old stellar population and recent star formation (SF), if there is no significant difference in their formation epoch between different spectral classes.
Non-passive red early-type galaxies (REGs) have larger metallicity and younger age than passive REGs, considering their UV -- optical -- NIR colours, which implies that non-passive REGs have suffered recent SF adding young and metal-rich stars to them. The radio detection fraction of REGs strongly depends on their optical absolute magnitudes, while that of most late-type galaxies does not, implying the difference in their radio sources: AGN and SF.
The optical -- NIR colours of red late-type galaxies (RLGs) reveal that they may have considerable old stars as well as young stars. The UV -- optical colours and the radio detection fraction of passive RLGs show that they have properties similar to REGs rather than non-passive RLGs. Dust extinction may not be a dominant factor making RLGs red, because RLGs are detected in the mid- and far-infrared bands less efficiently than blue late-type galaxies (BLGs). The passive BLGs have very blue UV -- optical -- NIR colours, implying either recent SF quenching or current SF in their outskirts.
Including star formation rate, other multi-wavelength properties in each fine class are investigated, and their implication on the identity of each fine class is discussed.
\end{abstract}


\section{Introduction}

One of the fundamental issues of the observational cosmology is the evolutionary connection between various classes of galaxies.
Since galaxies show a large range of physical properties and activities, they are classified using various criteria: morphology \citep[e.g.][]{par05}, colour \citep[e.g.][]{mar07}, spectral features \citep[e.g.][]{mat06}, and so on.
Knowledge about those various kinds of galaxies has been accumulated for a long time, motivating many efforts to find the connections between different galaxy classes and their implication on galaxy evolution.
However, since the galaxy classifications in most previous studies were limited to only one or two properties, some detailed aspects in galaxy evolution have not been sufficiently inspected.
For example, galaxies with unusual features, such as blue early-type galaxies \citep{fer05,lee06} or passive spiral galaxies \citep{yam04,ish07} are difficult to understand well in studies using simple galaxy classifications.
The necessity of fine classifications of galaxies was also pointed out in \citet{lee07}, who reported that the fundamental planes of early-type galaxies with different spectral classes have a potential difference in the slope, indicating that even morphologically similar galaxies have different natures if they have different spectral features.
Recently, \citet[][hereafter Paper I]{lee08} presented the optical properties of the Sloan Digital Sky Survey \citep[SDSS;][]{yor00} galaxies in various fine classes based on their morphology, colour and spectral features, showing that each galaxy class has its own distinguishable features and that an analysis based on a simple classification often has a risk of mixing up different kinds of objects with different nature.

This is the second in a series of papers on the nature of galaxies in the finely-divided classes. This paper is focused on the multi-wavelength properties of galaxies in the fine classes.
Since the source and mechanism of light emission are often different for different wavelength bands, multi-wavelength data provide much information about galaxies that is difficult to acquire from optical data only.
We divide the extra-optical bands into five categories: X-ray; ultraviolet (UV); near-infrared (NIR); mid- and far-infrared (MIR and FIR); and radio.

The X-ray is sensitive to hot thermal gas \citep{pie00,cox06,die07} or active galactic nucleus (AGN) activity \citep{app00,str05,and07}.
The UV light is sensitive to young stellar populations or very old stellar populations like core-helium burning
low mass stars \citep{yi99,dor03,nef05}. Since the UV $-$ optical colour is more sensitive to young stars than the optical colour, the UV information is useful to constrain the mean stellar ages of galaxies \citep{yi99,dor03}.
The NIR light is sensitive to the old stellar population and metal abundance in a galaxy \citep{sma01}. Since the optical light is sensitive both to the age and metallicity of galaxies, the combination of the information in the optical and NIR bands is useful to estimate the mean stellar age and mean stellar metallicity of galaxies \citep{sma01,cha06,lee08b}.
The MIR and FIR light is sensitive to the dust contents in galaxies \citep{dra01,bos04,xil04}. Since vigorous star formation (SF) is usually accompanied by rich dust production \citep{cal00,for04,kon04}, the information in the MIR and FIR bands is helpful to inspect SF activity. In addition, the MIR and FIR light is also sensitive to AGN activity \citep{fad02,haa04,bra06}.
The radio shows hot gas contents in a galaxy by means of supernovae-induced synchrotron radiation, which are direct probes to current SF activity free from dust extinction \citep{con92,cra98,kim01}. Moreover, since the surroundings of a central blackhole emit a large amount of synchrotron radio flux in some kinds of AGNs, the radio is also valuable in studies of the AGN activity in galaxies \citep{con02,odo02,cro07}.

Since multi-wavelength data provide advantages in probing the nature of galaxies, some previous studies used multiple data sets covering different wavelength bands.
\citet{cha06} estimated age, metallicity and $\alpha$-enhancement of 2728 elliptical galaxies using the SDSS and the Two Micron All Sky Survey \citep[2MASS;][]{jar00} data, finding that the optical colours of elliptical galaxies are sensitive to age, metallicity and $\alpha$-enhancement, while the optical $-$ NIR colours are sensitive to metallicity and to $\alpha$-enhancement, but are less sensitive to age.
\citet{got05} investigated the optical properties of 4248 infrared galaxies using the SDSS and the InfraRed Astronomical Satellite \citep[IRAS;][]{neu84} data, showing that the infrared luminosity and the optically estimated SF rate for star-forming galaxies have a good correlation.
\citet{ive01,ive02} analysed the optical properties of radio galaxies using the SDSS and the Faint Images of the Radio Sky at Twenty-centimetres \citep[FIRST;][]{bec95} data, reporting that radio galaxies have a different optical luminosity distribution from that of non-radio galaxies selected by the same criteria.
\citet{yi05} estimated the SF in early-type galaxies using the SDSS and the Galaxy Evolution Explorer \citep[GALEX;][]{mar03} data, presenting that the UV colour-magnitude relation (CMR) of early-type galaxies shows a substantially larger scatter than the optical CMR, which is evidence of recent SF activity.
\citet{and07} inspected the X-ray AGNs using the SDSS and the Roentgen Satellite \citep[ROSAT;][]{asc81} data, providing an expanded catalogue of 7000 AGNs.
\citet{obr06} conducted a very rare study on galaxy evolution using data covering all available wavelength bands, finding that there are strong correlations between the optical properties of galaxies and their detection fraction at other wavelengths.

These previous studies produced many important results about galaxy evolution, but much more still remains to be investigated. Multi-wavelength data sets provide very useful information to understand the nature of galaxies divided into fine classes, constraining their star formation history, obscured AGN activities, dust contents, ans so on.
We have been doing a comprehensive study on a set of fine galaxy classes in the SDSS, based on their morphology, colour and spectral features.
In this paper, the second in the series, we present the multi-wavelength properties of galaxies in various fine classes.
The outline of this paper is as follows.
Section 2 describes the data set we used, and \S3 briefly describes the methods to classify the SDSS galaxies and to select volume-limited samples.
We present the NIR properties of galaxies in each fine class in \S4. Section 5 displays the MIR and FIR properties, and radio properties are shown in \S6. UV properties and X-ray properties are presented in \S7 and \S8, respectively. In \S9, we estimate the star formation rate (SFR) of galaxies in each fine class using multiple methods.
Based on those multi-wavelength properties, we discuss the nature of galaxies in the fine classes in \S10. Finally, the conclusions in this paper are given in \S11.
Throughout this paper, we adopt the cosmological parameters 
$h=0.7$, $\Omega_{\Lambda}=0.7$, and $\Omega_{M}=0.3$.

\section{Data}

\subsection{Sloan Digital Sky Survey}

We use the SDSS Data Release 4 \citep[DR4;][]{ade06} \footnote{See http://www.sdss.org/dr4/.} main galaxy sample.
The SDSS is a photometric and spectroscopic survey, that has mapped about one quarter of the whole sky. The photometric and spectroscopic observations were conducted using the 2.5-m SDSS telescope at the Apache Point Observatory in New Mexico, USA.
The SDSS photometry was conducted in the $ugriz$ band images, with effective wavelengths of 3551, 4686, 6166, 7480 and 8932 {\AA}, respectively \citep{gun98}.
The astrometric uncertainty is less than 0.1$''$ at $r_{pet}<20.5$ \citep{pie03}.
The SDSS DR4 imaging data cover about 6670 deg$^{2}$, and the SDSS DR4 spectroscopic data cover 4783 deg$^{2}$. The median FWHM of point sources in the $r$ band images is $1.4''$, and the wavelength coverage in the spectroscopy is 3800 -- 9200{\AA}.
We use the photometric and structural parameters from the SDSS pipeline \citep{sto02} data, and the spectroscopic parameters from the Max-Planck-Institute for Astronomy catalogue \citep[MPA catalogue;][]{kau03,tre04,gal06}.
In addition, we use the velocity dispersion estimated using an automated spectroscopic pipeline called idlspec2d version 5 (D. Schlegel et al., in preparation).
Using the SDSS atlas images, we have measured the colour gradient, inverse concentration index, and the axis ratio of galaxies corrected for the inclination and the seeing effects \citep{par05,cho07}.

After foreground extinction correction \citep{sch98}, the magnitude of each galaxy was corrected in two aspects: the redshift effect (K-correction) and galaxy luminosity evolution (evolutionary correction). We used the method of \citet{bla03} to conduct K-correction, and the empirical formula of \citet{teg04} to conduct evolutionary correction. The evolutionary correction formula of \citet{teg04} was derived by means of a simple fit of galaxy luminosity evolution without galaxy type segregation, which means that it is not very accurate. However, since the evolutionary correction values in the redshift range of our galaxy sample are not large, we used the simple method of \citet{teg04}.
Using these methods, we corrected the observed magnitudes of about 360,000 SDSS galaxies into the magnitudes at redshift z = 0.1, where the SDSS galaxies were observed most frequently.
Since the corrections are applied optionally in this paper, we denote the magnitude with K-correction as $^{0.1{\textrm{\protect\scriptsize K}}}m$ and the magnitude with both K-correction and evolutionary correction as $^{0.1{\textrm{\protect\scriptsize KE}}}m$, if the observed magnitude is $m$. We use Petrosian magnitudes to represent optical brightness, while we use model magnitudes to calculate galaxy colours.
The SDSS official website\footnote{http://www.sdss.org/dr5/algorithms/photometry.html} recommends Petrosian magnitudes for flux estimation of bright (enough to be included in the SDSS spectroscopic sample) extended sources, while it recommends model magnitudes for colour calculation of extended sources, because the model magnitude is measured consistently in all bands, based on the best-fit parameters in the $r$ band.

\subsection{Two Micron All Sky Survey}

We use the 2MASS Extended Source Catalogue (XSC) \footnote{See http://www.ipac.caltech.edu/2mass/.}.
Using the Mt. Hopkins northern 1.3-m telescope and the CTIO southern 1.3-m telescope in Chile, the 2MASS covered almost entire sky in the $J(1.25\mu$m), $H(1.65\mu$m) and $K_{s}(2.17\mu$m) bands, with spatial resolution of $3''$. 
Point sources were detected, brighter than about 1 mJy at the $10\sigma$ limit, which corresponds to 16.3, 16.5 and 16.2 AB magnitude in the $J$, $H$ and $K_{s}$ bands, respectively. The magnitude limit of typical extended sources in the $K_{s}$ band is about 15.3 ABmag.
The astrometric uncertainty is less than $0.2''$.
The 2MASS objects were matched to the SDSS objects with $1.5''$ tolerance\footnote{In matching most data sets, we adopt the tolerance suggested by \citet{obr06}, except for the IRAS.}. If two or more 2MASS objects are found within $1.5''$ from an SDSS object, the nearest one was selected as the NIR counterpart of the SDSS object. Using this method, about 200,000 2MASS extended sources were matched with our SDSS sample.
We estimated the probability of false match by calculating the match number of the SDSS objects to random catalogue with the same number density as the 2MASS catalogue \citep{obr06}, finding that the false match probability between SDSS and 2MASS is about 0.005 per cent.
To estimate the optical $-$ NIR colour within the same aperture, we calculated the SDSS magnitude within the 2MASS fiducial circular aperture, using the method of \citet{cha06}. The magnitudes and colours in the 2MASS bands were corrected for the redshift effect using the method of \citet{bla03}. 

\subsection{InfraRed Astronomical Satellite}

We use the IRAS Faint Source Catalogue \citep[FSC;][]{mos92} \footnote{See http://irsa.ipac.caltech.edu/IRASdocs/surveys/fsc.html.}. The IRAS covered 98 per cent of all sky in the $12\mu$m, $25\mu$m, $60\mu$m and $100\mu$m bands, with spatial resolution between $1-5'$. The IRAS FSC catalogue contains about 173,000 sources, and the detection limit of the IRAS is of order of 1 Jy at 100 $\mu$m.
The IRAS FSC sources were matched to the SDSS objects within three times of the elliptical positional uncertainties of IRAS sources, using the method of \citet{hwa07}. The number of IRAS sources matched with our SDSS sample is about 7,000, and the false match probability between SDSS and IRAS is about 0.071 per cent.

\subsection{Faint Images of the Radio Sky at Twenty-Centimetres}

We use the FIRST data \footnote{See http://sundog.stsci.edu/.}. The FIRST survey covers over 10,000 square-degrees in the 1.4 GHz (20 cm) band, with spatial resolution of $5''$, and will cover a quarter of the sky matched to the SDSS coverage in the future. The completeness limit of the FIRST survey is 1 mJy, including about 1 million sources. The FIRST has the highest spatial resolution and the smallest positional errors among the large radio surveys.
The FIRST sources were matched to the SDSS objects with $3''$ tolerance, which returns about 14,000 sources with 0.45 per cent false match probability.

\subsection{NRAO VLA Sky Survey}\label{nvss}

We use the NRAO VLA Sky Survey \citep[NVSS;][]{con98} data \footnote{See http://www.cv.nrao.edu/nvss/.}, which covers the entire sky north of -40 deg declination in the 1.4 GHz (20 cm) band, with $45''$ resolution. The NVSS catalogue has about 2.5 mJy completeness limit, including about 1.8 million sources. The NVSS is complementary to the FIRST, because the FIRST sometimes splits an extended radio structure into multiple sources due to its very high resolution \citep{bec95,ive02,bes05a}.
The NVSS sources were matched to the SDSS objects with $15''$ tolerance, which returns about 11,000 sources with 9.40 per cent false match probability.

\subsection{Galaxy Evolution Explorer}

We use the GALEX survey GR2/GR3 data \footnote{See http://galex.stsci.edu/GR2/.}.
The GALEX was launched in 2003 April, and will cover the whole sky in the far-UV (FUV; 1350 -- 1750 \AA) and near-UV (NUV; 1750 -- 2800 \AA) bands, with spatial resolutions of $6.0''$ in the FUV and $4.5''$ in the NUV.
The GALEX sources were matched to the SDSS galaxies with $6''$ tolerance. In this paper, only the GALEX objects detected both in the NUV and FUV bands were used, the number of which is about 85,000. The false match probability between SDSS and GALEX is estimated to be about 1.46 per cent.
To calculate UV (and UV $-$ optical) colours, we used the AUTO magnitudes in the GALEX catalogue and the model magnitudes in the SDSS catalogue.

\subsection{Roentgen Satellite}

We use the ROSAT all sky survey data \footnote{See http://www.mpe.mpg.de/xray/wave/rosat/index.php.}.
The ROSAT included the X-ray Telescope (XRT) with a 2.4-m focal length mirror assembly, consisting of four nested Wolter-I mirrors. The focal plane instrumentation consisted of the Position Sensitive Proportional Counter (PSPC) and High Resolution Imager (HRI). The PSPC was used in the ROSAT all sky survey, which covered the all sky in the $0.1-2.4$ keV bands, with spatial resolution of $22''$ and with $2^{\circ}$ field of view in diameter.
About 100,000 sources were detected in the ROSAT survey.
We use the information of ROSAT sources matched with the SDSS objects, provided by the SDSS. In our sample, the number of ROSAT sources is about 2,200.

\section{Analysis}

\subsection{Classifications and sample volumes}

\begin{figure}
\includegraphics[width=84mm,height=84mm]{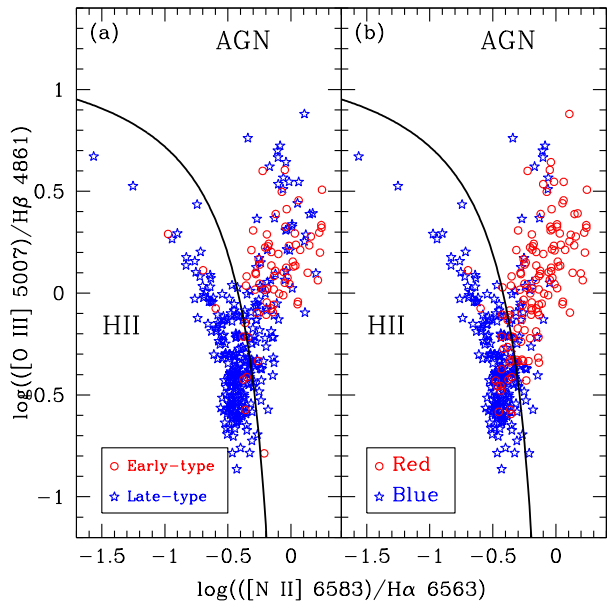}
\caption{ (a) Morphology and (b) colour class distributions on the BPT diagram for a small sample of the SDSS galaxies. In the panel (a), different symbols indicate different morphology: early-type galaxies (open circle) and late-type galaxies (open star). In the panel (b), different symbols indicate different colour: red galaxies (open circle) and blue galaxies (open star). The line shows the boundary dividing H{\protect\scriptsize II} galaxies and AGN host galaxies in this paper. }
\label{classify}
\end{figure}

\begin{table}
\centering
\caption{Abbreviations of the 16 fine galaxy classes}
\label{tclcl}
\begin{tabular}{lcccc}
\hline \hline
& \multicolumn{2}{c}{Early-type$^{(a)}$} & \multicolumn{2}{c}{Late-type$^{(a)}$} \\
& Red$^{(b)}$ & Blue$^{(b)}$ & Red$^{(b)}$ & Blue$^{(b)}$ \\
\hline
Passive$^{(c)}$ & $p$REG & $p$BEG & $p$RLG & $p$BLG \\
H{\protect\scriptsize II}$^{(c)}$ & $h$REG & $h$BEG & $h$RLG & $h$BLG \\
Seyfert$^{(c)}$ & $s$REG & $s$BEG & $s$RLG & $s$BLG \\
LINER$^{(c)}$ & $l$REG & $l$BEG & $l$RLG & $l$BLG \\
\hline \hline
\end{tabular}
\\
(a) Morphology classification: \citet{par05}. (b) Colour classification: \citet{lee06}; Paper I. (c) Spectra classification: \citet{kau03b,kew06}.
\end{table}

In this study, we classified the galaxies in the SDSS spectroscopic sample, using three criteria: morphology, colour and spectral features.
The following is a brief introduction of the galaxy classification scheme in this study, and more details are described in Paper I.
First, we divided the SDSS galaxies into early-type galaxies and late-type galaxies, using the method of \citet{par05} in the colour-colour gradient -- light-concentration parameter space. The completeness and reliability\footnote{The completeness is defined as the fraction of successfully-selected galaxies of a given class from the original sample using the classification scheme, while the reliability is defined as the
fraction of galaxies of the desired class from the selected sub-
sample \citep{par05}.}
reach about 90 per cent as claimed by \citet{par05}.
Second, the SDSS galaxies are classified into red galaxies and blue galaxies, using the method of \citet{lee06} based on the colour distribution of early-type galaxies as a function of redshift.
Third, based on the flux ratios between several spectral lines \citep[e.g. BPT diagram;][]{bal81,kau03b,kew06}, we classified the SDSS galaxies into passive galaxies, H{\protect\scriptsize II} galaxies, Seyfert galaxies and low ionisation nuclear emission region (LINER) galaxies.

Fig.~\ref{classify} shows the distributions of different morphology and colour classes on the BPT diagram. As mentioned previously, it is noted that the three kinds of classification do not always agree with each other, showing the necessity of fine classification.
The three galaxy classifications with different criteria return a total of 16 fine classes of galaxies: [early-type, late-type] $\times$ [red, blue] $\times$ [passive, H{\protect\scriptsize II}, Seyfert, LINER].
Hereafter, we use the following abbreviations for the 16 galaxy classes: REG (red early-type galaxy), BEG (blue early-type galaxy), RLG (red late-type galaxy), BLG (blue late-type galaxy), and $p$- (passive), $h$- (H{\protect\scriptsize II}), $s$- (Seyfert), $l$- (LINER). The abbreviations for 16 fine galaxy classes are listed in Table \ref{tclcl}.

\begin{table}
\centering
\caption{Three sample-volumes}
\label{volumes}
\begin{tabular}{cccc}
\hline \hline
$^{(a)}$& $^{0.1\textrm{\protect\scriptsize KE}}M_{\textrm{\protect\scriptsize pet}}(r)^{(b)}$ & Redshift$^{(c)}$ & Number of objects$^{(d)}$\\
\hline
V1 & $(-22.92:-20.71)$ & $(0.08 : 0.10)$ & 34198\\
V2 & $(-21.39:-19.19)$ & $(0.04 : 0.05)$ & 12452\\
V3 & $(-19.88:-18.07)$ & $(0.02 : 0.03)$ & 5491\\
\hline \hline
\end{tabular}
\\
(a) Volume name. (b) The range of $r$-band absolute magnitude after K-correction and evolutionary correction. (c) Redshift range. (d) The number of objects in the volume.
\end{table}

We selected three volume-limited samples in the luminosity versus redshift space.
Since the redshift dependence of galaxy properties at $z<0.1$ may not be significant, we made each volume have a small redshift range and a large luminosity range, adequate to investigate the luminosity dependence of galaxy properties.
Table \ref{volumes} summarises the luminosity ranges, redshift ranges and the numbers of objects in the three selected volumes (V1, V2 and V3). More details in the sample-volume selection were described in Paper I.

\subsection{Uncertainties in the classifications}

\subsubsection{Confidence of morphological classification}

It should be noted that the confidence level of morphological classification \citep{par05} does not hold for each fine class in this paper. In other words, the disagreement between the quantitatively-determined morphology and visually-determined morphology in some fine classes may be larger than that in the whole sample. Particularly, this disagreement will be found more easily for fainter objects, and this possible contamination may affect the statistical results in this paper.
For faint objects, however, the visual classification as well as the quantitative classification has inaccuracy, because the visual classification depends on human subjectivity. Thus, the visual classification is not always superior to quantitative classification, and the quantitatively-determined morphology itself is meaningful, although it is sometimes different from visually-determined morphology.
In short, it should be born in mind that the \emph{morphology class} in this paper does not always agree with visually-classified morphology, but the quantitative morphology used in this paper is worthy as itself.

\subsubsection{Signal-to-noise criteria in spectral classification}

\begin{table}
\centering
\caption{Spectral class fraction as a function of S/N criterion}
\label{spsn}
\begin{tabular}{ccccc}
\hline \hline
S/N & 2 & 3 & 5 & 10 \\
\hline
passive & $14.18\%$ & $20.55\%$ & $30.07\%$ & $42.19\%$	\\
H{\protect\scriptsize II} & $49.58\%$ & $47.06\%$ & $45.64\%$ & $46.07\%$	\\
Seyfert & $12.62\%$ & $11.49\%$ & $8.96\%$ & $5.24\%$ 	\\
LINER & $23.61\%$ & $20.70\%$ & $15.33\%$ & $6.50\%$ 	\\
\hline \hline
\end{tabular}
\end{table}

Table \ref{spsn} summarises the variation of spectral class fractions as a function of signal-to-noise ratio (S/N) criterion, showing that the spectral classification is sensitive to its S/N criterion. As the S/N criterion increases, the fraction of passive galaxies increases significantly and the fraction of AGN host galaxies decreases. The fractional variation of H{\protect\scriptsize II} galaxies is relatively small.
This shows that we should be careful when comparing statistical properties between different spectral classes, and should pay regard to the fact that there may be some contamination in each class.
However, contamination tends to make the properties of different fine classes seem to be less different, compared to the properties of \emph{genuinely-different} fine classes. In other words, the \emph{minimum} difference of properties between different spectral classes will be found, using our spectral classification.

\subsubsection{Aperture effect on spectral classification}

Suppose an emission line object with (S/N)$_{z=0.02}$ at z = 0.02. If we put this object at z = 0.1, the observed flux of the emission line will decrease into about S$_{z=0.1}$ = 0.036 S$_{z=0.02}$ in the cosmology adopted in this paper. If we consider only Poisson noise, the expected S/N is: (S/N)$_{z=0.1}$ = 0.19 (S/N)$_{z=0.02}$.
However, Paper I shows that this effect on the class fraction is not significant actually. If this effect is significant, clear discontinuities should be found in the class fraction between different volumes, but such features are hardly found in Fig.~6 of Paper I.
This is because the redshift effect is compensated by the \emph{aperture effect}.
The $3''$ fixed aperture in the SDSS spectroscopy covers different physical size at different redshift: 1.2 kpc at z $\sim0.02$, 2.9 kpc at z $\sim0.05$ and 5.5 kpc at z $\sim0.1$. If the spectral line emission is spatially evenly distributed in an object, the expected S/N variation ratio between z = 0.1 and z=0.02 is: $\big($(S/N)$_{z=0.1} \times$ (5.5 kpc)$\big)$ / $\big($(S/N)$_{z=0.02} \times$ (1.1 kpc)$\big)$ $\sim0.95$, which may not cause a significant systematic difference.

However, the problem is that the spectral line emission is not spatially even in real galaxies. For example, line emission from H{\protect\scriptsize II} region in late-type galaxies is biased to their outskirts, whereas line emission from an AGN concentrates on the centre of that galaxy. That is, a face-on late-type galaxy with a passive bulge of 1.5 kpc diameter at z=0.2 may be classified as a passive late-type galaxy in our classification scheme, in spite that it is not entirely passive. Therefore, it should be born in mind that some passive late-type galaxies in our sample may be face-on H{\protect\scriptsize II} late-type galaxies.

\section{Near Infrared}

\subsection{2MASS detection fraction and stellar mass} \label{sec2mass}

\begin{figure}
\includegraphics[width=84mm,height=110mm]{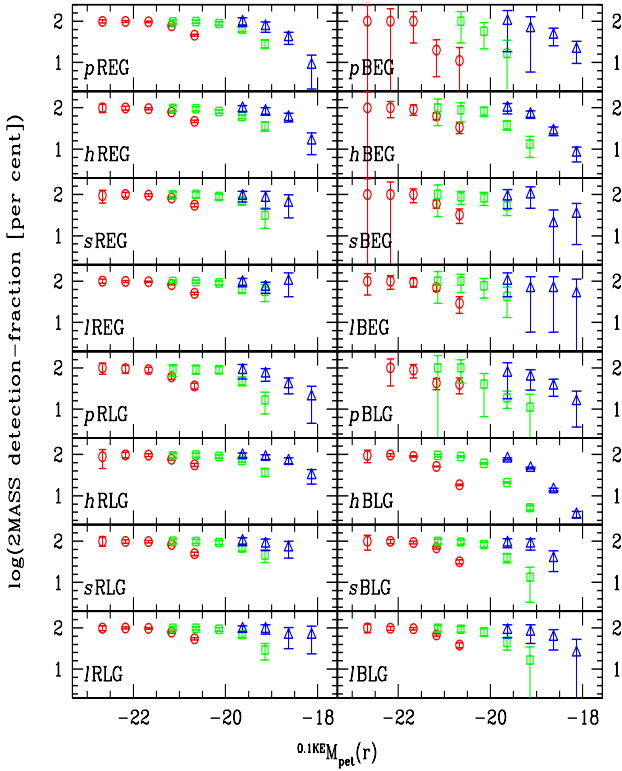}
\caption{The detection fraction of the SDSS objects in the 2MASS $K_{s}$ band. The open circles, open rectangles and open triangles represent the detection fractions (per cent) in V1, V2 and V3 volumes, respectively. Errorbars show the Poisson errors.}
\label{det2mass}
\end{figure}

\begin{table}
\centering
\caption{The number of the 2MASS-detected objects in the V1 volume\label{det2masstab}}
\begin{tabular}{lrr}
\hline \hline
& REG & BEG \\
\hline
Passive &  4102 (80.3 $\%$)  &	10 (38.5 $\%$)   \\
H{\protect\scriptsize II} &  1888 (76.2 $\%$)  &	93 (62.8 $\%$)    \\
Seyfert &  1200 (81.2 $\%$)  &	58 (59.2 $\%$)   \\
LINER & 4082 (84.5 $\%$)  &	117 (73.6 $\%$)    \\
\hline \hline
 & RLG & BLG \\
\hline
Passive  &	286 (64.1 $\%$)  &	43 (55.8 $\%$)  \\
H{\protect\scriptsize II}  &	1242 (76.4 $\%$)  &	4681 (48.6 $\%$)  \\
Seyfert &	1472 (83.2 $\%$)  &	1041 (70.7 $\%$)  \\
LINER   &	3036 (85.1 $\%$)  &	985 (76.9 $\%$)  \\
\hline \hline
\end{tabular}
\medskip
\\The percentages in the parentheses show the 2MASS detection fractions of the SDSS galaxies in the individual classes.
\end{table}

Table \ref{det2masstab} summarises the 2MASS detection fraction of the SDSS galaxies in each class, and Fig.~\ref{det2mass} shows the dependence of the 2MASS detection fraction on optical luminosity.
The discerned decreasing 2MAS detection fraction with decreasing luminosity is likely due to the difference in survey depths between the SDSS and the 2MASS.
\citet{obr06} showed that the SDSS galaxies detected by the 2MASS is essentially complete (completeness of $>$ 99 per cent) at $r_{pet}<16.3$, and \citet{mci06} showed that the 2MASS detects 90 per cent of the SDSS main galaxy sample brighter than $r_{pet}=17$. In our sample, the 2MASS detection fraction of the SDSS galaxies at $r_{pet}\le17$ is 92 per cent, which is consistent with the previous reports.

There are notable differences in the trend of the 2MASS detection fraction between different classes in Fig.~\ref{det2mass} and Table \ref{det2masstab}.
For example, the 2MASS detection fraction of blue galaxies decreases more rapidly as luminosity decreases than that of red galaxies does, which makes the detection fraction of the entire blue galaxies smaller than that of the entire red galaxies.
$p$BEGs are detected least in the 2MASS (39 per cent), and the detection fraction of $h$BLGs is also small (49 per cent). The difference in the 2MASS detection fraction between red galaxies and blue galaxies, is consistent with the optical colour dependence of 2MASS completeness shown by \citet{obr06}.
These trends are as expected, because typical young stars are relatively faint in the NIR band, while old stars are relatively bright in the NIR band.

\citet{obr06} reported that 2/3 of the SDSS galaxies with an AGN are detected by 2MASS, while only 1/10 of the SDSS star-forming galaxies are detected by 2MASS.
Since the data versions, selection criteria and classification methods in \citet{obr06} and in this paper are quite different, the detection fractions in the two studies do not agree with each other. However, a similar trend to that reported in \citet{obr06} is found in our BLGs, in the sense that faint $h$BLGs are detected less efficiently than faint AGN host BLGs (three bottom-right panels in Fig.~\ref{det2mass}). This trend is not clearly found in the other morphology-colour classes.
This difference between BLGs and the other classes may be due to the combination of the two continuum sources in the NIR band: old stellar population and AGN. 
Not only old stellar population \citep{sma01} but also an AGN contributes to the infrared continuum with its power-law spectral energy distribution (SED) \citep{buc06}. Since the fraction of old stellar population in BLGs is not large, the role of AGNs is relatively conspicuous in BLGs, as shown in Table \ref{det2masstab}.

\begin{figure}
\includegraphics[width=84mm,height=84mm]{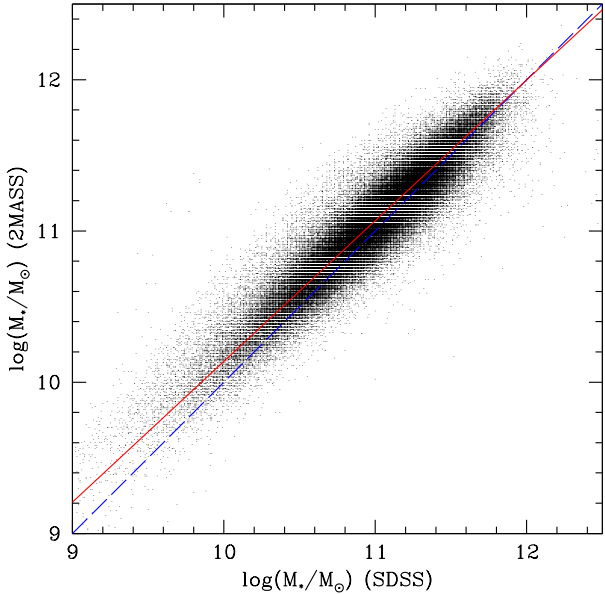}
\caption{ Stellar mass estimated using the SDSS \citep{gal06} versus stellar mass estimated using the 2MASS. The dashed line shows one to one relation, and the solid line presents the linear least-squares fit using the OLS bisector method \citep{iso90}. }
\label{stmass}
\end{figure}

\begin{table}
\centering
\caption{Median stellar mass of each fine class in the V1 volume\label{stmasstab}}
\begin{tabular}{lcc}
\hline \hline
& $\log$ (M$_{*}$/M$_{\odot}$) & $\log$ (M$_{*}$/M$_{\odot}$) \\
& using 2MASS & using SDSS$^{(a)}$ \\
\hline
$p$REG & $11.07\pm0.15$ & $10.90\pm0.17$ \\
$h$REG & $11.01\pm0.12$ & $10.87\pm0.15$ \\
$s$REG & $11.03\pm0.12$ & $10.93\pm0.15$ \\
$l$REG & $11.07\pm0.12$ & $10.99\pm0.15$ \\
\hline
$p$BEG & $11.09\pm0.16$ & $10.48\pm0.15$ \\
$h$BEG & $10.97\pm0.11$ & $10.67\pm0.18$ \\
$s$BEG & $10.95\pm0.11$ & $10.66\pm0.12$ \\
$l$BEG & $11.02\pm0.12$ & $10.70\pm0.18$ \\
\hline
$p$RLG & $11.09\pm0.15$ & $10.84\pm0.20$ \\
$h$RLG & $11.03\pm0.12$ & $10.98\pm0.16$ \\
$s$RLG & $11.07\pm0.12$ & $11.03\pm0.17$ \\
$l$RLG & $11.09\pm0.14$ & $11.06\pm0.18$ \\
\hline
$p$BLG & $10.96\pm0.14$ & $10.80\pm0.20$ \\
$h$BLG & $10.93\pm0.11$ & $10.70\pm0.17$ \\
$s$BLG & $10.99\pm0.12$ & $10.85\pm0.16$ \\
$l$BLG & $11.02\pm0.12$ & $10.91\pm0.16$ \\
\hline \hline
\end{tabular}
\medskip
\\Median($\log$ mass) $\pm$ SIQR($\log$ mass). (a) Estimated from the results of \citet{gal06} based on the SDSS Petrosian magnitudes \citep{gal06}.
\end{table}

We estimated the median stellar mass of the galaxies in each fine class, based on their absolute magnitudes in the $K_{s}$ band.
As the solar $K_{s}$ absolute magnitude, $5.14$ mag was adopted \citep{bla07}. Fig.~\ref{stmass} compares the 2MASS-based stellar mass with the stellar mass estimated using the SDSS Petrosian magnitude by \citet{gal06}.
In Fig.~\ref{stmass}, the two kinds of stellar mass of massive galaxies show good agreement, but the 2MASS stellar mass tends to be larger than the SDSS stellar mass as the mass decreases. Using the ordinary least squares (OLS) bisector method \citep{iso90}, the linear fit of the relation between the two masses is estimated as:
\begin{equation} \label{massrel}
\log(\textrm{M}_{*,\textrm{\scriptsize 2MASS}})
\\=(0.93\pm0.00) \times \log(\textrm{M}_{*,\textrm{\scriptsize SDSS}}) +0.84.
\end{equation}
Table \ref{stmasstab} summarises the 2MASS-based and the SDSS-based median stellar masses of each fine class.

\subsection{Colour-magnitude relation}

\begin{figure}
\includegraphics[width=84mm,height=110mm]{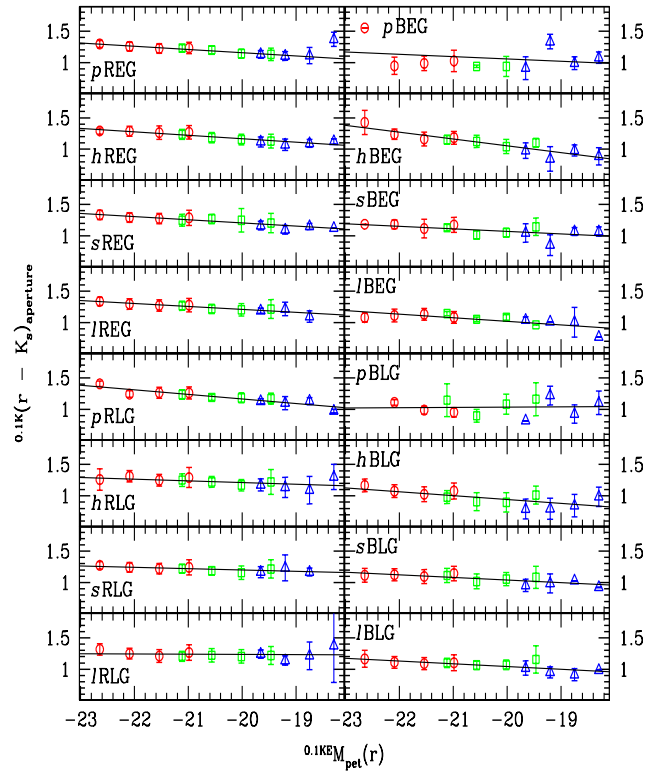}
\caption{$^{0.1\textrm{\protect\scriptsize K}}(r-K_{s})$ colour variation with respect to $^{0.1\textrm{\protect\scriptsize KE}}M_{\textrm{\protect\scriptsize pet}}$, for each class. Open circles show the median $^{0.1\textrm{\protect\scriptsize K}}(r-K_{s})$ colours at given magnitude bin in V1, and open rectangles and open triangles do in V2 and V3, respectively. Errorbars represent the sample inter-quartile ranges (SIQRs) of $^{0.1\textrm{\protect\scriptsize K}}(r-K_{s})$ colour at given magnitude bin. The solid line in each panel is the linear least-squares fit. For late-type galaxies, the axis ratio limit ($>0.6$) is applied.}
\label{Lrk}
\end{figure}

\begin{table}
\centering
\caption{The linear fits in the $^{0.1\textrm{\protect\scriptsize K}}(r-K_{s})$ versus $^{0.1\textrm{\protect\scriptsize KE}}M_{\textrm{\protect\scriptsize pet}}$ plots
\label{Lrkfit}}
\begin{tabular}{lcc}
\hline \hline
& REG & BEG \\
\hline
Passive & $-0.053\pm0.003(1.210)$	&	$-0.037\pm0.049(1.093)$	\\
H{\protect\scriptsize II} & $-0.055\pm0.007(1.220)$	&	$-0.108\pm0.015(1.163)$	\\
Seyfert & $-0.050\pm0.007(1.257)$	&	$-0.041\pm0.018(1.113)$	\\
LINER & $-0.047\pm0.007(1.256)$	&	$-0.056\pm0.015(1.075)$	\\
\hline \hline
& RLG & BLG \\
\hline
Passive & $-0.072\pm0.010(1.235)$	&	$0.004\pm0.035(1.031)$	\\
H{\protect\scriptsize II} & $-0.027\pm0.014(1.240)$	&	$-0.062\pm0.020(1.003)$	\\
Seyfert & $-0.020\pm0.007(1.221)$	&	$-0.042\pm0.010(1.081)$	\\
LINER & $-0.003\pm0.014(1.240)$	&	$-0.043\pm0.012(1.088)$	\\
\hline \hline
\end{tabular}
\medskip
\\The slopes of the median $^{0.1\textrm{\protect\scriptsize K}}(r-K_{s})$ with respect to $^{0.1\textrm{\protect\scriptsize KE}}M_{\textrm{\protect\scriptsize pet}}$, and $^{0.1\textrm{\protect\scriptsize K}}(r-K_{s})$ at $^{0.1\textrm{\protect\scriptsize KE}}M_{\textrm{\protect\scriptsize pet}}(r)=-21$ within parentheses.
\end{table}

Fig.~\ref{Lrk} shows the $^{0.1\textrm{\protect\scriptsize K}}(r-K_{s})$ properties for each class, and the linear fits in Fig.~\ref{Lrk} are summarised in Table \ref{Lrkfit}. The NIR magnitude is known to be less sensitive to dust extinction than optical magnitude. However, since the $^{0.1\textrm{\protect\scriptsize K}}r$ is an optical magnitude sensitive to dust extinction, the axis ratio limit \citep[$>0.6$;][]{cho07} was applied to late-type galaxies, to reduce the effect of the internal extinction.

Like the optical colour and magnitude, the $^{0.1\textrm{\protect\scriptsize K}}(r-K_{s})$ colour and $^{0.1\textrm{\protect\scriptsize KE}}M_{\textrm{\protect\scriptsize pet}}$ show good correlations in most classes.
The CMR slopes of REGs are similar regardless of their spectral classes.
\citet{cha06} estimated the $(r-K_{s})$ CMR slope of elliptical galaxies to be $-0.0731$, which is more negative than the $^{0.1\textrm{\protect\scriptsize K}}(r-K_{s})$ CMR slope of REGs in this study, $-0.055 - -0.047$. This may be mainly due to the difference in the criteria to select sample galaxies.
In other words, both REGs and BEGs may be included in the elliptical galaxy sample of \citet{cha06}.
$h$BEGs show a very rapid CMR slope in the $^{0.1\textrm{\protect\scriptsize K}}(r-K_{s})$ colour ($-0.11\pm0.02$):
the $^{0.1\textrm{\protect\scriptsize K}}(r-K_{s})$ CMR slope of $h$BEGs is twice as rapid as that of $h$REGs, and their slope difference is significant by $3.5\sigma$.

\begin{figure}
\includegraphics[width=84mm,height=84mm]{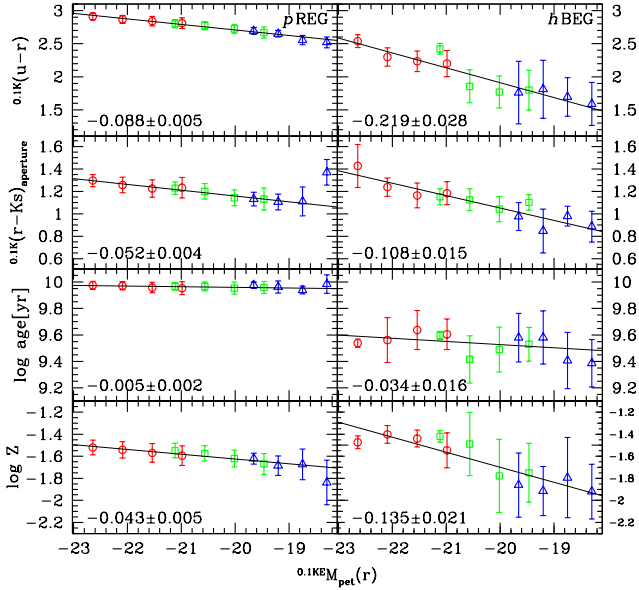}
\caption{The variations of $^{0.1\textrm{\protect\scriptsize K}}(u-r)$, $^{0.1\textrm{\protect\scriptsize K}}(r-K_{s})$, mean stellar age \citep{gal06} and mean stellar metallicity \citep{gal06} with respect to $^{0.1\textrm{\protect\scriptsize KE}}M_{\textrm{\protect\scriptsize pet}}$, for $p$REGs and $h$BEGs. Open circles show the median values of each quantity at give magnitude bin in V1, and open rectangles and open triangles do in V2 and V3, respectively. Errorbars represent the SIQR at given magnitude bin. The solid line in each panel is the linear least-squares fit and the fit slope is denoted at lower-left corner in each panel.}
\label{agemet}
\end{figure}

To check the effects of age and metallicity on the CMR slope of these early-type galaxies, we use the mean stellar age and metallicity information derived by \citet{gal06}.
Fig.~\ref{agemet} compares the optical magnitude dependence of optical and NIR colours, age and metallicity of $p$REGs and $h$BEGs. This confirms the previous knowledge that the CMR slope of typical red elliptical galaxies (i.e. $p$REGs) mainly originates from the luminosity dependence of their metallicity \citep[e.g.][]{kod97,kau98}.
On the other hand, $h$BEGs shows a larger spread in their age and metallicity than $p$REGs. The ages of $h$BEGs are much smaller than those of $p$REGs as expected, and the age -- magnitude relation of $h$BEGs is not linear. The median metallicity of bright $h$BEGs is larger than that of bright $p$REGs, whereas the median metallicity of faint $h$BEGs is smaller than that of faint $p$REGs. The age and metallicity of $h$BEGs as a function of absolute magnitude hint to a possible bimodality, with a potential dividing magnitude $^{0.1\textrm{\protect\scriptsize KE}}M(r)\sim-20.5$. However, the scatter (particularly for faint $h$BEGs) is too large to draw any more robust conclusions.
Both the ages and the metallicities may affect the rapid $^{0.1\textrm{\protect\scriptsize K}}(r-K_{s})$ CMR slope of $h$BEGs in Fig.~\ref{agemet}, but the effect of metallicity seem to be larger than that of age.

The CMR of $p$RLGs is more similar to those of REGs than those of non-passive RLGs. Since most $p$RLGs are expected to be bulge-dominated face-on galaxies (Paper I), this result is understood to be due to the similarity between elliptical galaxies and spiral bulges \citep{jab07}.
It is interesting that non-passive RLGs have small negative CMR slopes in Fig.~\ref{Lrk}.
Particularly, the $^{0.1\textrm{\protect\scriptsize K}}(r-K_{s})$ CMR slopes of AGN host RLGs are very small (but still negative), compared to those in REGs, whereas AGN host (i.e. Seyfert or LINER) galaxies in other morphology-colour classes have similar CMR slopes to those of REGs. This trend is not found in the $^{0.1\textrm{\protect\scriptsize K}}(u-r)$ CMR (Paper I).
Why AGN host RLGs (and why AGN host non-RLGs do not) have such unusual $^{0.1\textrm{\protect\scriptsize K}}(r-K_{s})$ CMR slopes, is an open question.

In Paper I, $^{0.1\textrm{\protect\scriptsize K}}(u-r)$ CMR slope of $h$BLGs is significantly more negative than that of REGs. On the other hand, the $^{0.1\textrm{\protect\scriptsize K}}(r-K_{s})$ CMR slope of $h$BLGs is similar to that of REGs.
Since $^{0.1\textrm{\protect\scriptsize K}}(r-K_{s})$ is less sensitive to stellar age than $^{0.1\textrm{\protect\scriptsize K}}(u-r)$ \citep{sma01}, the CMR slope variation of $h$BLGs between the optical and NIR bands supports the conclusion in Paper I that the rapid CMR slope of $h$BLGs in the optical band is strongly affected by age difference. In other words, the faint $h$BLGs are younger than the bright $h$BLGs.

\subsection{Colour-colour relation\label{S2massccr}}

\begin{figure*}
\includegraphics[width=168mm]{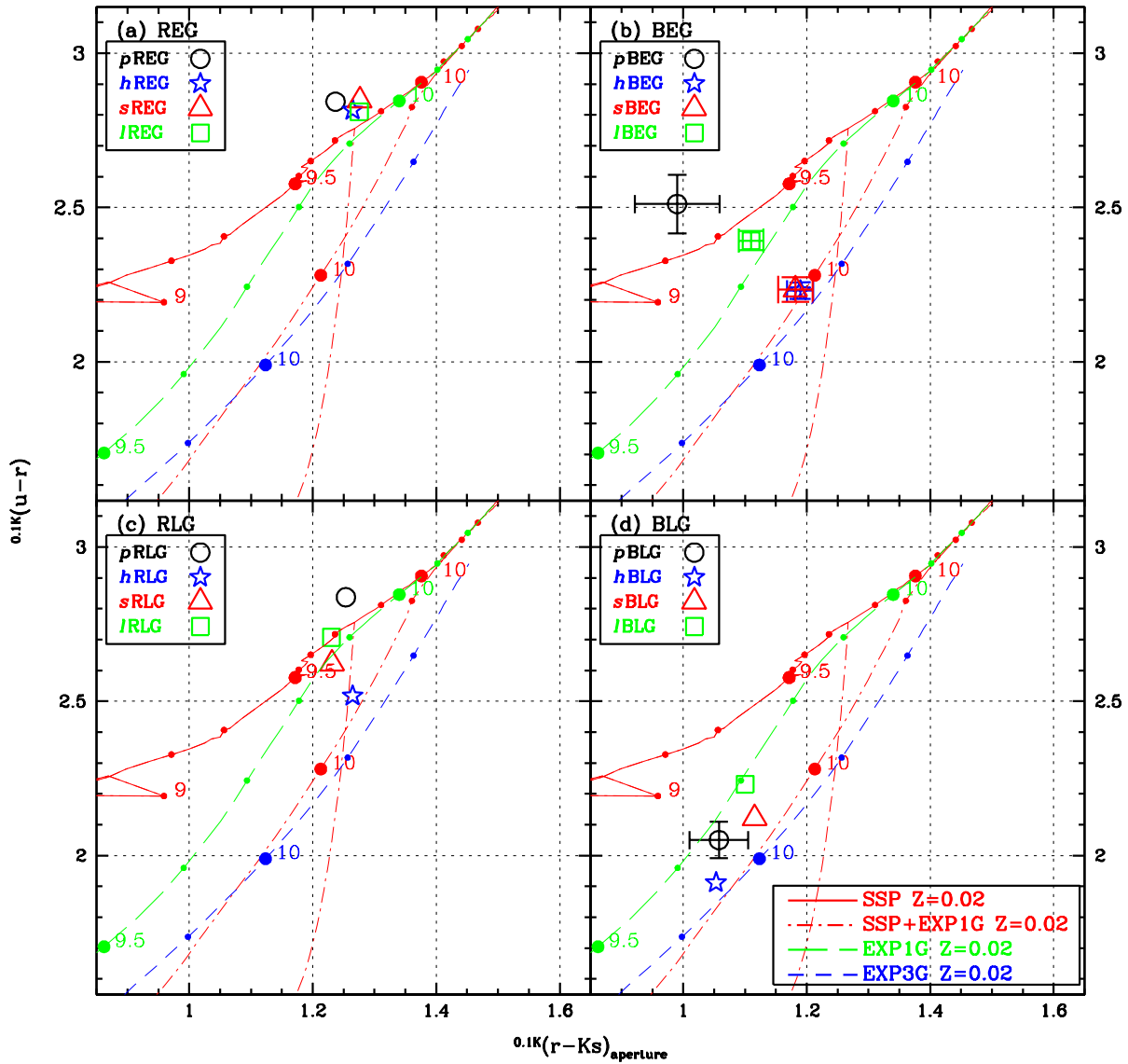}
\caption{ The $^{0.1\textrm{\protect\scriptsize K}}(u-r)$ versus $^{0.1\textrm{\protect\scriptsize K}}(r-K_{s})$ relation of (a) REGs, (b) BEGs, (c) RLGs and (d) BLGs in the V1 volume.
Each symbol represents the median colours of each galaxy class, and the errorbars represent their sampling errors. The sampling errors for the classes without errorbars are smaller than the symbol size. The median colours of each class is compared to the population synthesis models of \citet{bru03}.
The solid line shows the SSP model \citep{bru03} with Z=0.02, and the dot-dashed line displays the 90 per cent SSP + 10 per cent secondary star formation (SF) in an exponentially decreasing SF history (EXP) with the exponential time-scale of 1 Gyr since 7 Gyr old.
The long-dashed line shows the EXP model with Z=0.02 and the time-scale of 1 Gyr.
The short-dashed line is the same as the green line, except that the time-scale is 3 Gyr. 
On the model lines, it is dotted at every 0.1 dex of age, and the numbers beside filled circles represent the log scales of the age in year since the first SF. }
\label{optnir1}
\end{figure*}

\begin{table}
\centering
\caption{The median SDSS and 2MASS colours of each class in the V1 volume\label{optnir1tab}}
\begin{tabular}{rcc}
\hline \hline
 & $^{0.1\textrm{\protect\scriptsize K}}(r-K_{s})$ & $^{0.1\textrm{\protect\scriptsize K}}(u-r)$ \\
\hline
$p$REG  &  1.237$\pm$0.003  &  2.843$\pm$0.003	\\
$h$REG  &  1.265$\pm$0.005  &  2.816$\pm$0.005	\\
$s$REG  &  1.276$\pm$0.005  &  2.811$\pm$0.006	\\
$l$REG  &  1.276$\pm$0.003  &  2.846$\pm$0.003	\\
\hline
$p$BEG  &  0.990$\pm$0.069  &  2.511$\pm$0.095	\\
$h$BEG  &  1.190$\pm$0.021  &  2.231$\pm$0.027	\\
$s$BEG  &  1.182$\pm$0.028  &  2.235$\pm$0.040	\\
$l$BEG  &  1.110$\pm$0.020  &  2.393$\pm$0.031	\\
\hline
$p$RLG  &  1.254$\pm$0.013  &  2.837$\pm$0.014	\\
$h$RLG  &  1.265$\pm$0.009  &  2.518$\pm$0.009	\\
$s$RLG  &  1.231$\pm$0.006  &  2.622$\pm$0.007	\\
$l$RLG  &  1.230$\pm$0.004  &  2.707$\pm$0.005	\\
\hline
$p$BLG  &  1.058$\pm$0.048  &  2.050$\pm$0.059	\\
$h$BLG  &  1.053$\pm$0.004  &  1.912$\pm$0.005	\\
$s$BLG  &  1.116$\pm$0.008  &  2.120$\pm$0.011	\\
$l$BLG  &  1.101$\pm$0.007  &  2.231$\pm$0.010	\\
\hline \hline
\end{tabular}
\medskip
\\The $\pm$ values are the sampling errors of the median colour.
\end{table}

Fig.~\ref{optnir1} shows the median loci of the galaxy classes on the $^{0.1\textrm{\protect\scriptsize K}}(u-r)$ versus $^{0.1\textrm{\protect\scriptsize K}}(r-K_{s})$ diagram, in the V1 volume.
Errors were estimated by calculating the standard deviation of median values in 200-times repetitive random sampling. The median colours and sampling errors in Fig.~\ref{optnir1} are listed in Table \ref{optnir1tab}. For late-type galaxies, the axis ratio limit ($>0.6$) was applied.

REGs (Fig.~\ref{optnir1}a) show narrowly-ranged median colours about $^{0.1\textrm{\protect\scriptsize K}}(u-r)\sim2.85$ and $^{0.1\textrm{\protect\scriptsize K}}(r-K_{s})\sim1.25$, which is consistent with the $6-8$ Gyr old simple stellar population (SSP) galaxies modeled by \citet{bru03}. It is noted that $p$REGs have slightly bluer $^{0.1\textrm{\protect\scriptsize K}}(r-K_{s})$ colour than the other REGs.
$p$RLGs have comparable colour to REGs, and RLGs (Fig.~\ref{optnir1}c) are located on a sequence from REG colours to decreasing $^{0.1\textrm{\protect\scriptsize K}}(u-r)$ in the diagram.
This sequence is roughly consistent with the model of old SSP + recent (or current) SF in Fig.~\ref{optnir1}. Since RLGs have typically red centre and blue outskirts (Paper I), the colour sequence suggests that RLGs have large bulges with similar stellar contents to those of REGs and discs with current SF.

BEGs (Fig.~\ref{optnir1}b) show two different loci in the colour-colour diagram. $p$BEGs and $l$BEGs have colours consistent with those of the $2-3$ Gyr old SSP model galaxies. On the other hand, supposed that there was no secondary burst, $h$BEGs and $s$BEGs have consistent colours with those of the $11-12$ Gyr old `exponentially-decreasing SF rate with 3 Gyr time-scale (EXP3G)' model galaxies.
However, since the estimated ages ($11-12$ Gyr) of $h$BEGs and $s$BEGs based on the EXP3G model are unreasonably old (even older than REGs), $h$BEGs and $s$BEGs probably consist of old stars and very recently formed stars. Adopting the SSP+EXP1G model in Fig.~\ref{optnir1}, $h$BEGs and $s$BEGs are estimated to be younger than 10 Gyr old and seem to have had secondary starburst very recently.
It is noted that the direction to which the model galaxy colour evolves depends on the strength and duration of the second burst. For example, if the second burst is SSP, the model galaxy colour exactly reverses the original SSP evolutionary track, which means that the formation epoch of $p$BEGs and $l$BEGs may be much older than $2-3$ Gyr.
\citet{fer05} investigated the colour evolution of blue early-type galaxies at z $\sim0.7$ in the \emph{HST} fields, showing that their colours are well explained if they had secondary starburst events. The BEGs in our sample seem to be the nearby counterparts of those intermediate-z objects.

BLGs (Fig.~\ref{optnir1}d) are bluer than BEGs in the $^{0.1\textrm{\protect\scriptsize K}}(u-r)$ colour, but BLGs have the $^{0.1\textrm{\protect\scriptsize K}}(r-K_{s})$ colour range similar to BEGs. Since it is known that the SF history of typical blue spiral galaxies is complicated, it is not easy to infer the exact evolutionary path of BLGs in the diagram. However, it is clear that BLGs have younger stellar populations than any other morphology-colour class, on average.
It is notable that $p$BLGs seem to have very young mean stellar age, in spite that they are spectroscopically passive. This implies that these galaxies may be passive just within the SDSS spectroscopy fibre aperture ($3''$; 4.5 kpc at z = 0.08 and 5.5 kpc at z = 0.1), or that these galaxies may have become passive recently.
We checked the 2MASS-detected $p$BLGs in the V1 volume by eye, finding that at least 12 of 43 (about 28 per cent) $p$BLGs seem to have passive central regions but have star-forming spiral arms in their outskirts.

\begin{table}
\centering
\caption{The median optical -- NIR colours in different volumes\label{optnirvar}}
\begin{tabular}{ccccccc}
\hline \hline
 & \multicolumn{3}{c}{$^{0.1\textrm{\protect\scriptsize K}}(r-K_{s})$} & \multicolumn{3}{c}{$^{0.1\textrm{\protect\scriptsize K}}(u-r)$} \\
 & V1 & V2 & V3 & V1 & V2 & V3 \\
\hline
$p$REG  &  1.24 & 1.19 & 1.13 & 2.84 & 2.76 & 2.65	\\
$h$REG  &  1.26 & 1.18 & 1.09 & 2.82 & 2.72 & 2.59	\\
$s$REG  &  1.28 & 1.25 & 1.12 & 2.81 & 2.73 & 2.62	\\
$l$REG  &  1.28 & 1.24 & 1.18 & 2.85 & 2.78 & 2.65	\\
\hline
$p$BEG  &  0.99 & 0.94 & 1.03 & 2.51 & 2.34 & 2.38	\\
$h$BEG  &  1.19 & 1.08 & 0.96 & 2.23 & 1.93 & 1.80	\\
$s$BEG  &  1.18 & 1.05 & 0.99 & 2.24 & 2.22 & 2.13	\\
$l$BEG  &  1.11 & 1.07 & 1.00 & 2.39 & 2.37 & 2.37	\\
\hline
$p$RLG  &  1.25 & 1.18 & 1.09 & 2.84 & 2.76 & 2.61	\\
$h$RLG  &  1.27 & 1.21 & 1.16 & 2.52 & 2.56 & 2.56	\\
$s$RLG  &  1.23 & 1.19 & 1.15 & 2.62 & 2.58 & 2.55	\\
$l$RLG  &  1.23 & 1.20 & 1.21 & 2.71 & 2.67 & 2.64	\\
\hline
$p$BLG  &  1.06 & 1.11 & 0.92 & 2.05 & 2.16 & 2.33	\\
$h$BLG  &  1.05 & 0.93 & 0.80 & 1.91 & 1.78 & 1.64	\\
$s$BLG  &  1.12 & 1.06 & 0.95 & 2.12 & 2.08 & 2.04	\\
$l$BLG  &  1.10 & 1.07 & 0.97 & 2.23 & 2.18 & 2.30	\\
\hline \hline
\end{tabular}
\end{table}

We found that the optical and NIR median colours of the galaxies in most classes become bluer as the sample volume varies: V1 $\rightarrow$ V2 $\rightarrow$ V3, as listed in Table \ref{optnirvar}.
This indicates that the mean stellar age of galaxies in each class becomes younger as the variation of the volumes from V1 to V3.
Since each volume has different luminosity and redshift ranges, the variation of mean stellar age reflects the luminosity or redshift dependence of mean stellar age. However, because it is not plausible that nearby galaxies (e.g. V3 galaxies) are younger than  distant galaxies (e.g. V1 galaxies) at fixed luminosity, the age difference between the volumes are probably due to the luminosity dependence of mean stellar age of galaxies.
That is, the fainter galaxies in a given class are younger than the brighter galaxies, which is consistent with the galaxy downsizing scenario \citep{cow96,ber05,tre05}.

\section{Mid and Far Infrared}

\subsection{IRAS detection fraction and IR luminosity}
\begin{figure}
\includegraphics[width=84mm,height=110mm]{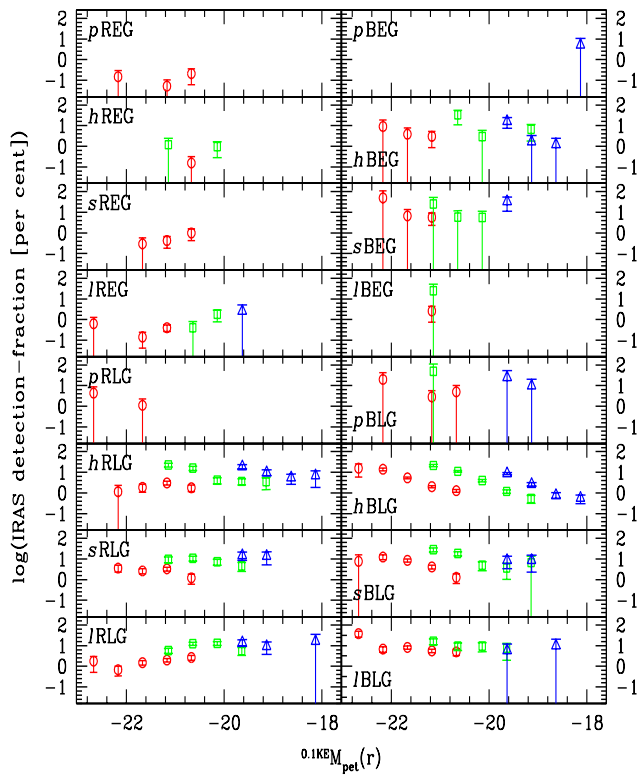}
\caption{The detection fraction of the SDSS objects in the IRAS. The symbols are the same as those in Fig.~\ref{det2mass}.}
\label{detiras}
\end{figure}

\begin{table}
\centering
\caption{The number of the IRAS-detected objects in the V1 volume
\label{detirastab}}
\begin{tabular}{lrr}
\hline \hline
 & REG & BEG \\
\hline
Passive &    4 (0.08 $\%$)  & 0 (0.00 $\%$)  \\
H{\protect\scriptsize II} &    1 (0.04 $\%$)  & 4 (2.70 $\%$)   \\ 
Seyfert &    7 (0.47 $\%$)  & 5 (5.10 $\%$)    \\
LINER &   11 (0.23 $\%$)  & 2 (1.26 $\%$)   \\
\hline \hline
 & RLG & BLG \\
\hline
Passive  & 2 (0.45 $\%$)  & 3 (3.90 $\%$) \\
H{\protect\scriptsize II}   & 38 (2.34 $\%$)  & 267 (2.77 $\%$) \\ 
Seyfert   & 48 (2.71 $\%$)  & 81 (5.50 $\%$)  \\
LINER   & 61 (1.71 $\%$)  & 94 (7.34 $\%$)  \\
\hline \hline
\end{tabular}
\medskip
\\The percentages in the parentheses show the IRAS detection fractions of the SDSS galaxies in the individual classes.
\end{table}

Fig.~\ref{detiras} shows the optical luminosity dependence of the IRAS detection fractions in each galaxy class, and Table \ref{detirastab} lists the total IRAS detection fractions in each galaxy class.
As shown in Fig.~\ref{detiras} and Table \ref{detirastab}, REGs are poorly detected in the IRAS, compared to other morphology-colour classes. This is because the major source of the MIR and FIR emission is dust, which is typically accompanied by SF \citep{kon04}.
Table \ref{detirastab} confirms the results of \citet{obr06}, which reported a strong bias of the IRAS detection towards optically blue galaxies. Moreover, Table \ref{detirastab} shows that morphology as well as colour is an important factor affecting the IRAS detection fraction, in the sense that late-type galaxies are detected more efficiently than early-type galaxies in the IRAS.
Among early-type galaxies, $h$BEGs and $s$BEGs show relatively large detection fractions (up to more than ten per cent at given magnitude bin), but the number of detected galaxies is very small (less than ten).
We visually checked the IRAS-detected early-type galaxies in the V1 volume, finding that 7 of 23 REGs and 3 of 11 BEGs (about 30 per cent of the IRAS-detected early-type galaxies) have very nearby neighbours and that one $l$BEG seems to be a misclassified $l$BLG. However, the other early-type galaxies do not show any unusual feature apparently, which may be dust-rich early-type galaxies.

According to \citet{obr06}, over 90 per cent of IRAS galaxies show strong emission lines in their optical spectra. Fig.~\ref{detiras} and Table \ref{detirastab} confirm those results of \citet{obr06}, noting that 99 per cent of the IRAS-detected objects are non-passive.
Among late-type galaxies, the detection fraction of BLGs is larger than that of RLGs. If most RLGs were dust-richer than BLGs, RLGs would be detected more efficiently than BLGs in the IRAS. Therefore, this result implies that dust extinction may not be a dominant factor making RLGs redder than BLGs and that another factor (such as bulge-to-disc ratio) may play an important role in their colour difference.

\begin{table}
\centering
\caption{IR luminosity of each fine class in the V1 volume
\label{lirtab}}
\begin{tabular}{lcc}
\hline \hline
 & $\log$(L(IR)$^{(a)}$/L$_{\odot}$) & $\log$(L(FIR)$^{(b)}$/L$_{\odot}$) \\
\hline
$p$REG & $11.48\pm0.03$ & $11.15\pm0.07$ \\  
$h$REG & $11.66\pm0.00$ & $11.47\pm0.00$ \\
$s$REG & $11.52\pm0.08$ & $11.25\pm0.09$ \\
$l$REG & $11.52\pm0.09$ & $11.14\pm0.13$ \\
\hline
$p$BEG & --- & --- \\
$h$BEG & $11.51\pm0.23$ & $11.33\pm0.12$ \\
$s$BEG & $11.71\pm0.02$ & $11.55\pm0.16$ \\
$l$BEG & $11.62\pm0.04$ & $11.33\pm0.06$ \\
\hline
$p$RLG & $11.52\pm0.06$ & $11.29\pm0.12$ \\
$h$RLG & $11.48\pm0.08$ & $11.18\pm0.11$ \\
$s$RLG & $11.45\pm0.09$ & $11.13\pm0.09$ \\
$l$RLG & $11.53\pm0.09$ & $11.22\pm0.12$ \\
\hline
$p$BLG & $11.40\pm0.04$ & $11.06\pm0.10$ \\
$h$BLG & $11.49\pm0.09$ & $11.19\pm0.09$ \\
$s$BLG & $11.54\pm0.06$ & $11.21\pm0.09$ \\
$l$BLG & $11.53\pm0.08$ & $11.24\pm0.11$ \\
\hline \hline
\end{tabular}
\medskip
\\Median($\log$ luminosity) $\pm$ SIQR($\log$ luminosity). (a) Median luminosity at 8 -- 1000 $\mu$m of the IRAS-detected galaxies. (b) Median luminosity at 40 -- 500 $\mu$m of the IRAS-detected galaxies.
\end{table}

We estimate the total and far infrared luminosity of our sample galaxies, using the conventional formulae in \citet{san96}.
Table \ref{lirtab} lists median IR luminosity of each fine class in the V1 volume. No significant difference in the median IR luminosity of the IRAS-detected galaxies is found between the fine classes, but it should be considered that each fine class has an IR-detection fraction significantly different from the others.
It is noted that the IR luminosity of BLGs is very similar to that of RLGs, indicating that dust in RLGs is not richer than that in BLGs.

\subsection{Colour-colour relation}

\begin{figure}
\includegraphics[width=84mm]{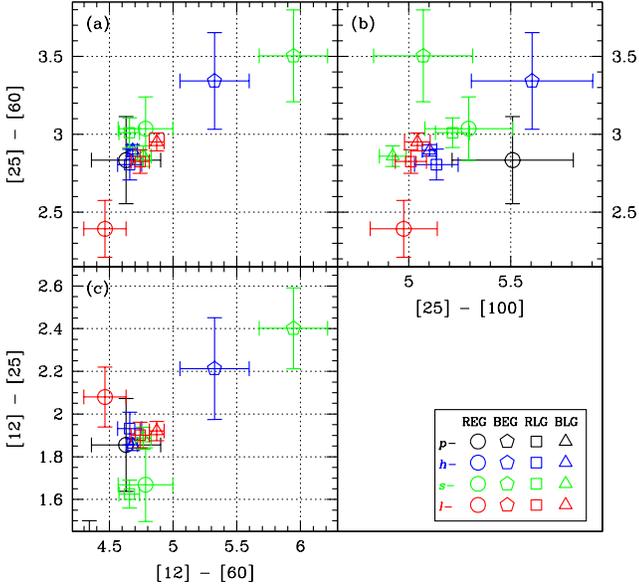}
\caption{ (a) The [12]$-$[60] versus [25]$-$[60] relation; (b) the [25]$-$[100] versus [25]$-$[60] relation; and (c) the [12]$-$[25] versus [25]$-$[60] relation of each class in the V1 volume. Five classes ($h$REGs, $p$BEGs, $l$BEGs, $p$RLGs and $p$BLGs) with the data point number fewer than four are not pointed on the plot, because their sampling errors are too large ($>0.3$). See the text for the definition of [12], [25], [60], and [100]. }
\label{irasccr1}
\end{figure}

\begin{table}
\centering
\caption{The median IRAS colours of each class in the V1 volume\label{irascolour}}
\begin{tabular}{rcccc}
\hline \hline
 & [12]$-$[25] & [12]$-$[60] & [25]$-$[60] & [25]$-$[100] \\
\hline
$p$REG  &  1.86$\pm$0.22 & 4.63$\pm$0.28 & 2.83$\pm$0.28 & 5.51$\pm$0.30	\\
$h$REG  &  2.40$\pm$0.50 & 4.63$\pm$0.58 & 2.23$\pm$0.50 & 5.75$\pm$0.57	\\
$s$REG  &  1.67$\pm$0.17 & 4.78$\pm$0.21 & 3.03$\pm$0.20 & 5.29$\pm$0.22	\\
$l$REG  &  2.08$\pm$0.14 & 4.47$\pm$0.17 & 2.39$\pm$0.18 & 4.98$\pm$0.16	\\
\hline
$p$BEG  &  --- & --- & --- & ---	\\
$h$BEG  &  2.21$\pm$0.24 & 5.32$\pm$0.27 & 3.34$\pm$0.31 & 5.60$\pm$0.30	\\
$s$BEG  &  2.40$\pm$0.19 & 5.94$\pm$0.27 & 3.50$\pm$0.30 & 5.07$\pm$0.24	\\
$l$BEG  &  2.05$\pm$0.32 & 5.42$\pm$0.34 & 3.70$\pm$0.41 & 5.08$\pm$0.41	\\
\hline
$p$RLG  &  1.41$\pm$0.33 & 4.37$\pm$0.37 & 3.01$\pm$0.39 & 6.07$\pm$0.44	\\
$h$RLG  &  1.93$\pm$0.08 & 4.66$\pm$0.09 & 2.81$\pm$0.10 & 5.14$\pm$0.11	\\
$s$RLG  &  1.63$\pm$0.07 & 4.66$\pm$0.08 & 3.01$\pm$0.10 & 5.21$\pm$0.08	\\
$l$RLG  &  1.90$\pm$0.06 & 4.74$\pm$0.07 & 2.83$\pm$0.08 & 5.01$\pm$0.08	\\
\hline
$p$BLG  &  1.22$\pm$0.28 & 4.34$\pm$0.30 & 3.00$\pm$0.34 & 5.38$\pm$0.32	\\
$h$BLG  &  1.86$\pm$0.03 & 4.68$\pm$0.04 & 2.90$\pm$0.04 & 5.10$\pm$0.04	\\
$s$BLG  &  1.89$\pm$0.05 & 4.77$\pm$0.06 & 2.86$\pm$0.07 & 4.92$\pm$0.07	\\
$l$BLG  &  1.92$\pm$0.05 & 4.87$\pm$0.06 & 2.95$\pm$0.06 & 5.04$\pm$0.06	\\
\hline \hline
\end{tabular}
\medskip
\\The $\pm$ values are the sampling errors of the median colour.
\end{table}

Fig.~\ref{irasccr1} shows the colour-colour diagrams using IRAS colours in the V1 volume. Each `IRAS magnitude' is defined as follows:
$[12] = 3.63 - 2.5\times \log(12\mu m$ flux);
$[25] = 2.07 - 2.5\times \log(25\mu m$ flux);
$[60] = 0.19 - 2.5\times \log(60\mu m$ flux);
$[100] = -0.92 - 2.5\times \log(100\mu m$ flux) \citep{wal88,wal89}.
The median IRAS colours and their sampling errors for each class used in Fig.~\ref{irasccr1} are summarised in Table \ref{irascolour}.

All galaxies detected in the IRAS have IR colours consistent with those of blue nebulae, Seyferts or quasars in \citet{wal89}.
It is noted that the $[12]-[60]$ colour and the $[25]-[60]$ colour show a tight relation in Fig.~\ref{irasccr1}a.
The tight IR colour-colour relation becomes scattered by substituting the $[25]-[100]$ colour for the $[12]-[60]$ colour in Fig.~\ref{irasccr1}b. However, the relation in Fig.~\ref{irasccr1}b is similar to that in Fig.~\ref{irasccr1}a, when a couple of classes with sampling errors larger than 0.25 ($s$BEGs, $h$BEGs and $p$REGs) are excluded.
In Fig.~\ref{irasccr1}c, the galaxy classes except for BEGs show a weak anti-correlation between the $[12]-[60]$ colour and the $[12]-[25]$ colour, implying that IRAS-detected BEGs have distinguishable IR SEDs from those of the other classes.
It is noted that the correlations between different IRAS colours in Fig.~\ref{irasccr1} are still found even if we confine our IRAS sample to the minimum source reliability $> 90$ per cent, but they are very difficult to find if the IRAS sample is limited to its minimum source reliability $> 95$ per cent.

Today, polycyclic aromatic hydrocarbons \citep[PAHs;][]{leg89} are regarded as one important factor determining SEDs in the IR bands. \citet{des07} showed that the IR flux ratio of 25 $\mu$m to 60 $\mu$m ($f_{\nu}(25)/f_{\nu}(60)$) have an anti-correlation to the strength of the PAH equivalent width (EW(PAH)), and that ultraluminous infrared galaxies with different spectral classes have different loci in the EW(PAH) versus $f_{\nu}(25)/f_{\nu}(60)$ diagram. That is, Seyfert galaxies have typically large $f_{\nu}(25)/f_{\nu}(60)$ and small EW(PAH), while H{\protect\scriptsize II} galaxies have typically small $f_{\nu}(25)/f_{\nu}(60)$ and large EW(PAH). LINER galaxies have loci in that diagram similar to those of H{\protect\scriptsize II} galaxies rather than those of Seyfert galaxies.
In Fig.~\ref{irasccr1}a, the distribution of the $[25]-[60]$ colours of the galaxy classes is partly consistent with the result of \citet{des07}, in the sense that Seyfert galaxies are redder than LINER galaxies on average.
However, the difference in the $[25]-[60]$ colour between Seyfert and H{\protect\scriptsize II} galaxies is very small in our result.
The colours of H{\protect\scriptsize II} galaxies are similar to those of Seyfert galaxies rather than those of LINER galaxies (Fig.~\ref{irasccr1}a,b), or intermediate between those of Seyfert galaxies and LINER galaxies (Fig.~\ref{irasccr1}c).
Since the sampling errors in Fig.~\ref{irasccr1} are so large that the statistical reliability is not high, these results need to be improved using a larger IR sample in the future.

\section{Radio}

\subsection{FIRST and NVSS detection fractions}

\begin{figure}
\includegraphics[width=84mm,height=110mm]{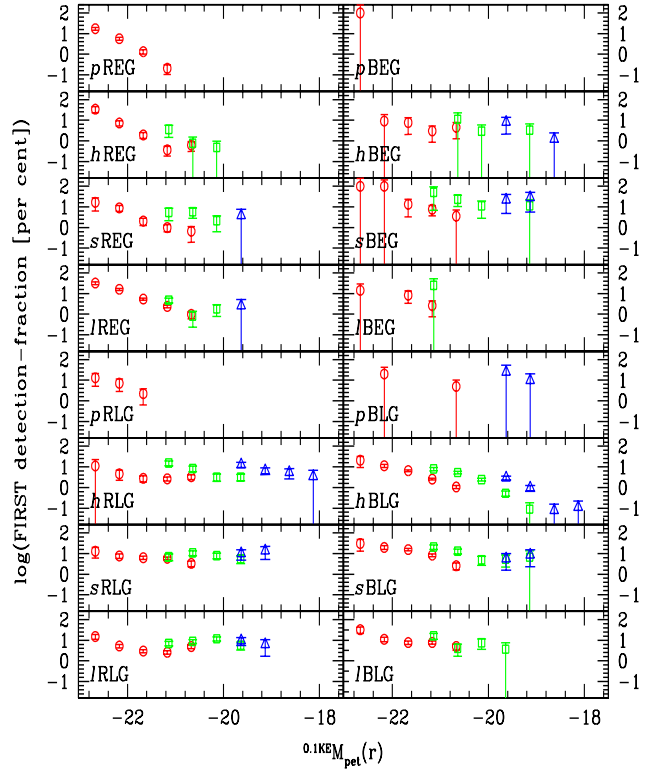}
\caption{The detection fraction of the SDSS objects in the FIRST. The symbols are the same as those in Fig.~\ref{det2mass}.}
\label{detfirst}
\end{figure}

\begin{figure}
\includegraphics[width=84mm,height=110mm]{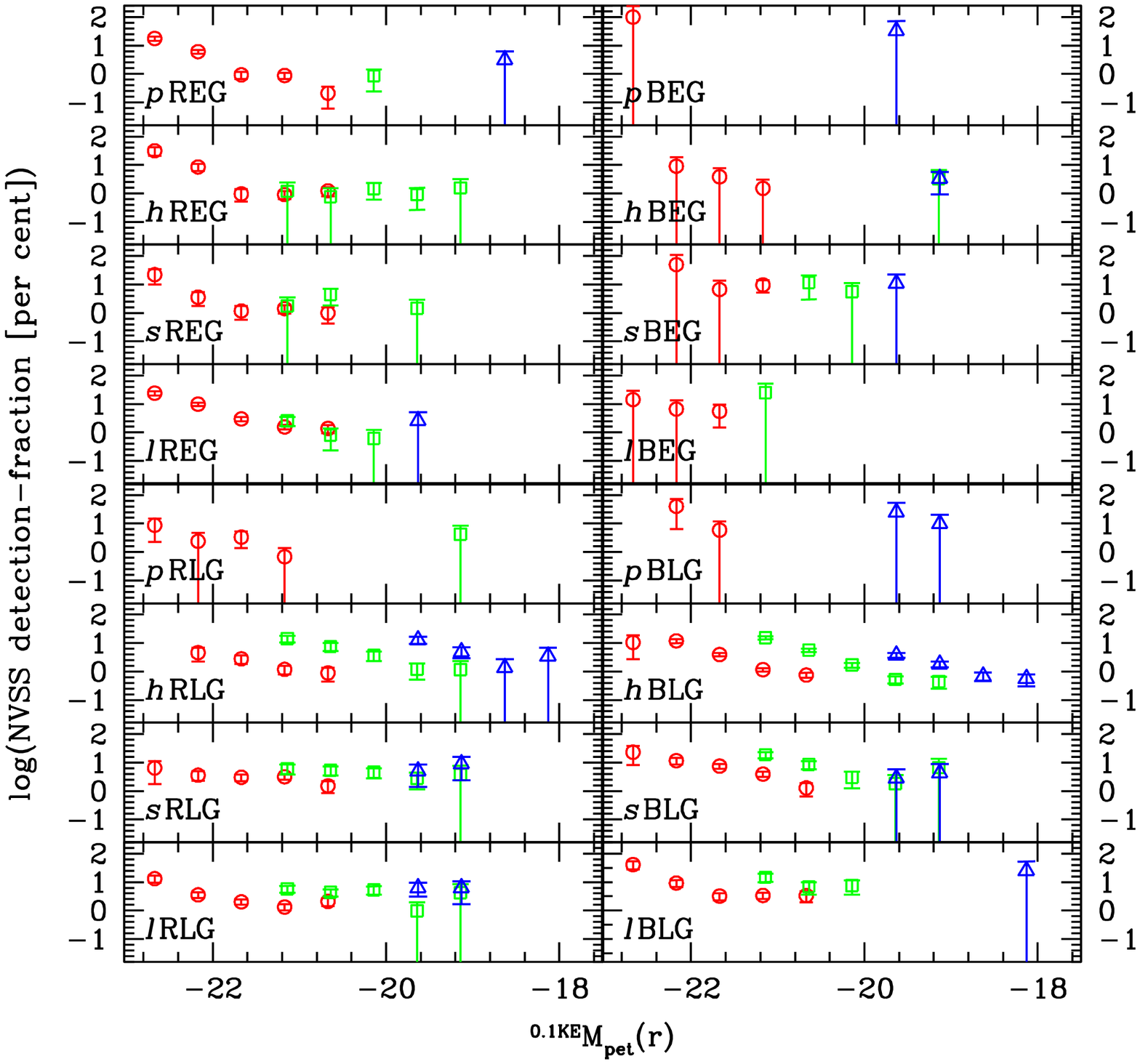}
\caption{The detection fraction of the SDSS objects in the NVSS. The symbols are the same as those in Fig.~\ref{det2mass}.}
\label{detnvss}
\end{figure}

\begin{table}
\centering
\caption{The number of the FIRST-detected objects in the V1 volume\label{detfirsttab}}
\begin{tabular}{lrrrr}
\hline \hline
& REG & BEG  \\
\hline
Passive &  103 (2.0 $\%$)   & 1 (3.8 $\%$)     \\
H{\protect\scriptsize II} &   44 (1.8 $\%$)   & 7 (4.7 $\%$)     \\
Seyfert &   29 (2.0 $\%$)   & 10 (10.2 $\%$)  \\ 
LINER &  272 (5.6 $\%$)   & 6 (3.8 $\%$)    \\
\hline \hline
& RLG & BLG \\
\hline
Passive   & 8 (1.8 $\%$)   & 2 (2.6 $\%$)   \\
H{\protect\scriptsize II}  & 49 (3.0 $\%$)   & 295 (3.1 $\%$)   \\
Seyfert  & 104 (5.9 $\%$)   & 152 (10.3 $\%$)  \\ 
LINER   & 139 (3.9 $\%$)   & 108 (8.4 $\%$)   \\
\hline \hline
\end{tabular}
\medskip
\\The percentages in the parentheses show the FIRST detection fractions of the SDSS galaxies in the individual classes.
\end{table}

\begin{table}
\centering
\caption{The number of the NVSS-detected objects in the V1 volume\label{detnvsstab}}
\begin{tabular}{lrrrr}
\hline \hline
& REG & BEG  \\
\hline
Passive &  117 (2.3 $\%$)   & 1 (3.8 $\%$)     \\
H{\protect\scriptsize II} &   50 (2.0 $\%$)   & 3 (2.0 $\%$)     \\
Seyfert &   25 (1.7 $\%$)   & 7 (7.1 $\%$)  \\ 
LINER &  177 (3.7 $\%$)   & 4 (2.5 $\%$)    \\
\hline \hline
& RLG & BLG \\
\hline
Passive   & 7 (1.6 $\%$)   & 3 (3.7 $\%$)   \\
H{\protect\scriptsize II}  & 26 (1.6 $\%$)   & 181 (1.9 $\%$)   \\
Seyfert  & 52 (2.9 $\%$)   & 77 (5.2 $\%$)  \\ 
LINER   & 86 (2.4 $\%$)   & 65 (5.1 $\%$)   \\
\hline \hline
\end{tabular}
\medskip
\\The percentages in the parentheses show the NVSS detection fractions of the SDSS galaxies in the individual classes.
\end{table}

The FIRST (NVSS) detection fraction of the SDSS galaxies in each class is listed in Table \ref{detfirsttab} (Table \ref{detnvsstab}), and Fig.~\ref{detfirst} (Fig.~\ref{detnvss}) displays the dependence of the FIRST (NVSS) detection fractions on optical luminosity. 
A difference in the radio detection trend is found between early-type galaxies and late-type galaxies. That is, the early-type galaxies with faint optical absolute magnitude are detected less efficiently in the radio than late-type galaxies with faint optical absolute magnitude.
\citet{bes05b} and \citet{cro07} showed that the fraction of radio-loud AGN host galaxies is a strong function of stellar mass, in the sense that radio-loud AGNs tend to be in massive galaxies. Thus, the rare radio-detection of the early-type galaxies with faint optical absolute magnitude implies that AGNs may be responsible for the radio emission of early-type galaxies.
On the other hand, the late-type galaxies with faint optical absolute magnitude are detected in the radio relatively well, implying that non-AGN activity (e.g. SF) may play an important role to make them bright in the radio.

It is interesting that the luminosity dependence of the radio detection fraction in $p$RLGs is similar to those of REGs, rather than to those of non-passive RLGs. Six of the nine FIRST-detected $p$RLGs are visually identified as face-on disc galaxies with bright bulges. Actually, those six galaxies are not easy to distinguish from elliptical galaxies by eye. Their relatively low light-concentrations indicate the existence of faint disc components in them, but they seem to be on the boundary between early-type galaxies and late-type galaxies. In other words, they would be classified as early-type galaxies if slightly more generous criteria were applied.
Two of the nine FIRST-detected $p$RLGs have very nearby satellite galaxies (overlapped with host galaxies on the images) that affect the light-concentrations of the host galaxies to be underestimated.
One of the nine FIRST-detected $p$RLGs is identified as an obvious (not-face-on) disc galaxy with a bright bulge.

Fig.~\ref{detfirst}--\ref{detnvss} and Table \ref{detfirsttab}--\ref{detnvsstab} show that blue galaxies are detected in the radio more efficiently than red galaxies, although the number itself of the radio-detected red galaxies is larger than that of the radio-detected blue galaxies.
AGN host galaxies (particularly, Seyfert galaxies) show relatively large detection fractions in the radio, compared to the other classes.
It is noted that about 9 per cent of the radio-detected galaxies are optically \emph{passive}, which confirms the disagreement between \emph{optical activity} and \emph{radio activity} previously reported by \citet{bes05b}.

\subsection{Radio luminosity distribution}

\begin{figure}
\includegraphics[width=84mm]{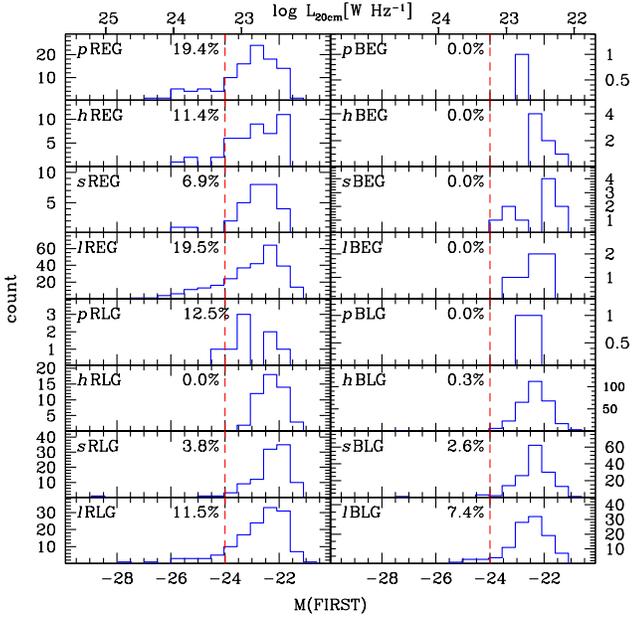}
\caption{ Radio (FIRST) luminosity distribution of each galaxy class in the V1 volume. The lower X-axis label shows absolute AB magnitude, and the upper X-axis label shows log-scale radio luminosity. The number fraction of galaxies with $M$(FIRST) $<-24$ is displayed at the upper-centre, and the $M$(FIRST) $=-24$ boundary is drawn as a vertical dashed-line in each panel. }
\label{firstlf}
\end{figure}

\begin{figure}
\includegraphics[width=84mm]{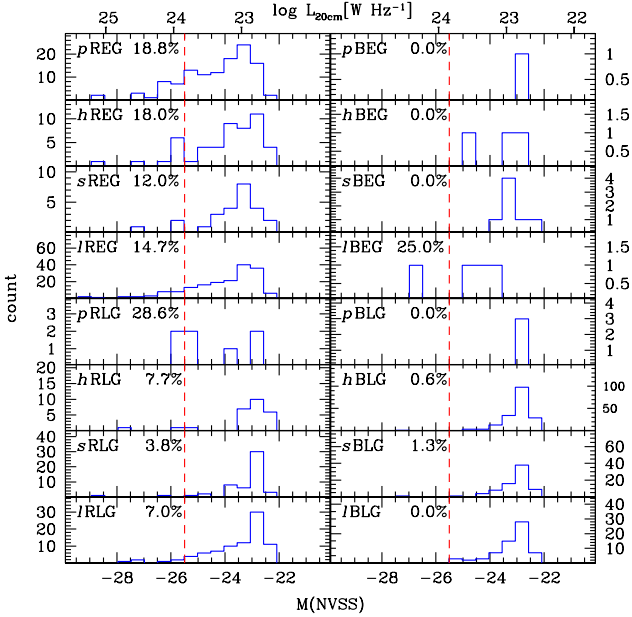}
\caption{ Radio (NVSS) luminosity distribution of each galaxy class in the V1 volume. The number fraction of galaxies with $M$(NVSS) $<-25.5$ is displayed at the upper-centre, and the $M$(NVSS) $=-25.5$ boundary is drawn as a vertical dashed-line in each panel. }
\label{nvsslf}
\end{figure}

Fig.~\ref{firstlf} (Fig.~\ref{nvsslf}) shows the distribution of FIRST (NVSS) luminosity of galaxies matched with SDSS galaxies of each class in the V1 volume. The $M$(FIRST) ($M$(NVSS)) is the absolute AB magnitude of a FIRST (NVSS) source without K-correction.
In Fig.~\ref{firstlf}, REGs have relatively large fraction of radio-loud galaxies, compared to the other classes. Particularly, almost 20 per cent of $p$REGs and $l$REGs are \emph{radio-loud} (defined as $M$(FIRST) $<-24$ in this paper), while only 8.6 per cent of galaxies in the entire classes are \emph{radio-loud}.
It is noted that $p$RLGs, $l$RLGs and $l$BLGs have large fractions of \emph{radio-loud} galaxies, too.
In Fig.~\ref{nvsslf}, the NVSS sources are brighter than the FIRST sources by $\sim1.5$ mag on average, showing that there may be underestimation of radio flux in the FIRST (by splitting a single structure into multiple sources due to its higher resolution) or overestimation in the NVSS (by binding multiple sources into a single source due to its lower resolution).
While 8.5 per cent of galaxies in the entire classes are $M$(NVSS) $<-25.5$, the fraction of REGs and $p$RLGs with $M$(NVSS) $<-25.5$ is larger than 12 per cent. Although some details in the radio luminosity distribution are different between the FIRST and the NVSS, the overall trends are similar: REGs have a relatively large fraction of \emph{radio-loud} galaxies.

\citet{bes05a} derived the radio luminosity functions (LFs) of AGN radio sources and SF radio sources, revealing that the bright radio sources are mostly AGNs.
Fig.~\ref{firstlf} confirms the result of \citet{bes05a}, noting that 10.6 per cent of AGN host galaxies are \emph{radio-loud}, while only 1.5 per cent of H{\protect\scriptsize II} galaxies are \emph{radio-loud}. Particularly, 14.7 per cent of LINER galaxies are \emph{radio-loud}.
It is notable that a large fraction of passive red galaxies are  \emph{radio-loud} galaxies (18.9 per cent).
One simple explanation for those objects is that they may have obscured AGNs \citep{ale03,vol04}. However, according to recent studies, there is no evidence that radio-loud AGNs without emission line have obscuring tori, and the correlation between the radiative efficiency of accretion and the phase of intergalactic medium may be responsible for the disagreement between emission-line AGNs and radio galaxies \citep{eva06,har07,kau08}.

\section{Ultraviolet}

\subsection{GALEX detection fraction}

\begin{figure}
\includegraphics[width=84mm,height=110mm]{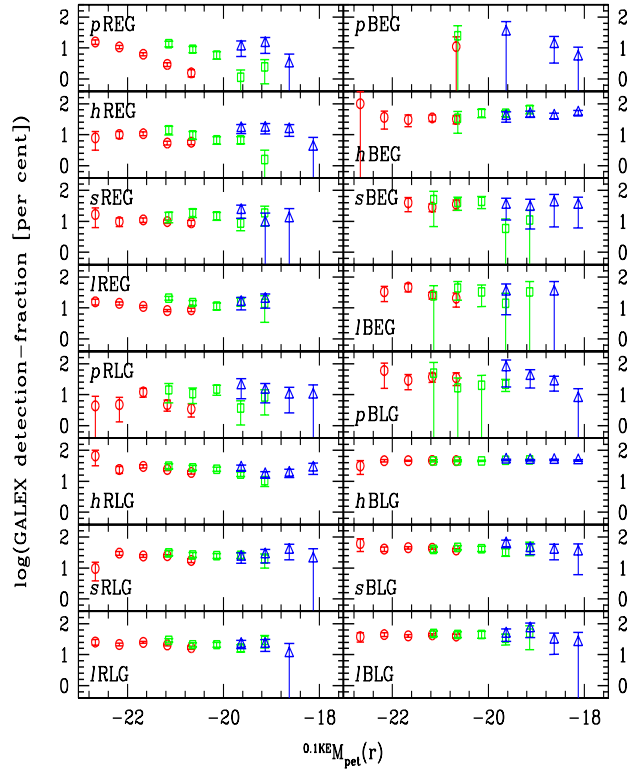}
\caption{The detection fraction of the SDSS objects in the GALEX. The symbols are the same as those in Fig.~\ref{det2mass}.}
\label{detgalex}
\end{figure}

\begin{table}
\centering
\caption{The number of the GALEX-detected objects in the V1 volume\label{detgalextab}}
\begin{tabular}{lrrrr}
\hline \hline
& REG & BEG  \\
\hline
Passive &  266 (5.2 $\%$)   & 1 (3.8 $\%$)    \\
H{\protect\scriptsize II} &  173 (7.0 $\%$)   & 50 (33.8 $\%$)   \\ 
Seyfert &  148 (10.0 $\%$)  & 31 (31.6 $\%$)    \\
LINER &  489 (10.1 $\%$)  & 47 (29.6 $\%$)   \\
\hline \hline
& RLG & BLG \\
\hline
Passive  & 26 (5.8 $\%$)   & 28 (36.4 $\%$)  \\
H{\protect\scriptsize II}  & 395 (24.3 $\%$)  & 4508 (46.8 $\%$)  \\ 
Seyfert  & 424 (24.0 $\%$)  & 630 (42.8 $\%$)  \\
LINER  & 777 (21.8 $\%$)  & 542 (42.3 $\%$)  \\
\hline \hline
\end{tabular}
\medskip
\\The percentages in the parentheses show the GALEX detection fractions of the SDSS galaxies in the individual classes.
\end{table}

Fig.~\ref{detgalex} presents the optical luminosity dependence of the GALEX detection fractions in each galaxy class, and the total GALEX detection fractions in each galaxy class are summarised in Table \ref{detgalextab}.
The detection fraction in the GALEX is second-largest among the multi-wavelength data sets used in this paper, next to the 2MASS.
\citet{obr06} showed that the SDSS galaxies detected in the GALEX are biased to blue galaxies. Consistent with that, the classes with the largest detection fraction in the GALEX are BLGs, about 40 per cent, while the classes with the smallest detection fraction are REGs: 10 per cent or less.
In a given morphology-colour class, non-passive galaxies are detected more efficiently than passive galaxies.
All of these trends indicate that young stellar populations are the major UV source.

\subsection{Colour-magnitude relation}

\begin{figure}
\includegraphics[width=84mm,height=110mm]{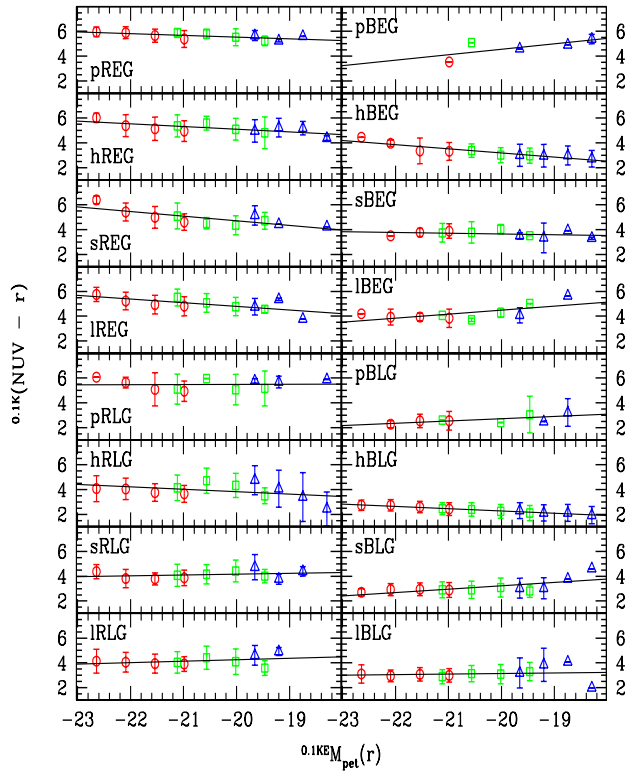}
\caption{ $^{0.1\textrm{\protect\scriptsize K}}($NUV$-r)$ colour variation with respect to $^{0.1\textrm{\protect\scriptsize KE}}M_{\textrm{\protect\scriptsize pet}}$. The symbols are the same as Fig.~\ref{Lrk}. }
\label{nuvr}
\end{figure}

\begin{table}
\centering
\caption{The linear fits in the $^{0.1\textrm{\protect\scriptsize K}}$(NUV$-r)$ versus $^{0.1\textrm{\protect\scriptsize KE}}M_{\textrm{\protect\scriptsize pet}}$ plots\label{nuvrfit}}
\begin{tabular}{lcc}
\hline \hline
& REG & BEG \\
\hline
Passive & $-0.140\pm0.049(5.685)$	&	$0.439\pm0.253(4.115)$	\\
H{\protect\scriptsize II} & $-0.219\pm0.066(5.317)$	&	$-0.335\pm0.048(3.521)$	\\
Seyfert & $-0.369\pm0.097(5.074)$	&	$-0.060\pm0.064(3.710)$	\\
LINER & $-0.293\pm0.102(5.088)$	&	$0.326\pm0.129(4.162)$	\\
\hline \hline
& RLG & BLG \\
\hline
Passive & $0.014\pm0.105(5.478)$	&	$0.181\pm0.079(2.538)$	\\
H{\protect\scriptsize II} & $-0.194\pm0.133(4.030)$	&	$-0.185\pm0.012(2.467)$	\\
Seyfert & $0.065\pm0.082(4.095)$	&	$0.273\pm0.090(2.959)$	\\
LINER & $0.114\pm0.109(4.137)$	&	$0.043\pm0.118(3.108)$	\\
\hline \hline
\end{tabular}
\medskip
\\The slopes of the median $^{0.1\textrm{\protect\scriptsize K}}$(NUV$-r)$ with respect to $^{0.1\textrm{\protect\scriptsize KE}}M_{\textrm{\protect\scriptsize pet}}$, and $^{0.1\textrm{\protect\scriptsize K}}$(NUV$-r)$ at $^{0.1\textrm{\protect\scriptsize KE}}M_{\textrm{\protect\scriptsize pet}}(r)=-21$ within parentheses.
\end{table}

Fig.~\ref{nuvr} shows the $^{0.1\textrm{\protect\scriptsize K}}$(NUV$-r)$ CMR in each class, and the linear fits in Fig.~\ref{nuvr} are listed in Table \ref{nuvrfit}. Since the UV magnitude is sensitive to dust extinction, the axis ratio limit ($>0.6$) was applied to late-type galaxies. 
In Fig.~\ref{nuvr}, $p$REGs show relatively tight $^{0.1\textrm{\protect\scriptsize K}}$(NUV$-r)$ CMRs.
The $^{0.1\textrm{\protect\scriptsize K}}$(NUV$-r)$ CMR slope of $p$REGs is $-0.14\pm0.05$, which is in agreement with the $^{0.1\textrm{\protect\scriptsize K}}$(NUV$-r)$ CMR slope, $-0.24\pm0.15$, of early-type galaxies at $0.05<z<0.10$, estimated by \citet{yi05}.
However, the $^{0.1\textrm{\protect\scriptsize K}}$(NUV$-r)$ CMRs of REGs are not as tight as the $^{0.1\textrm{\protect\scriptsize K}}(u-r)$ CMR of REGs shown in Paper I, suggesting that some REGs may have young stellar populations that hardly affect the optical colours, as pointed out by \citet{yi05}.
In addition, unlike $^{0.1\textrm{\protect\scriptsize K}}(u-r)$ CMRs of REGs that are similar between different spectral classes (Paper I), $^{0.1\textrm{\protect\scriptsize K}}$(NUV$-r)$ CMRs of REGs are different between different spectral classes.
The $^{0.1\textrm{\protect\scriptsize K}}$(NUV$-r)$ CMR slopes of non-passive REGs are more negative than that of $p$REGs, and the CMR scatters of non-passive REGs are larger than that of $p$REGs.
\citet{kav07} showed that the $^{0.1\textrm{\protect\scriptsize K}}$(NUV$-r)$ colour fluctuation and recent SF in some luminous early-type galaxies can be reproduced by numerical simulations of minor mergers.
According to the result of \citet{kav07}, the large scatters in $^{0.1\textrm{\protect\scriptsize K}}$(NUV$-r)$ CMR of non-passive REGs shows the possibility that many non-passive REGs have suffered recent minor merger events.
The difference in light concentration between $p$REGs and non-passive REGs partly supports this interpretation, in the sense that non-passive REGs are less concentrated than $p$REGs.
In Paper I, the median inverse concentration index of $p$REGs is $0.338\pm0.002$, which is smaller (more concentrated) than those of non-passive REGs ($0.342\pm0.001$, $0.347\pm0.002$ and $0.341\pm0.001$ for $h$REGs, $s$REGs and $l$REGs, respectively).

In Fig.~\ref{nuvr}, the CMRs of most late-type galaxy classes except for $h$BLGs are not tight.
The loose $^{0.1\textrm{\protect\scriptsize K}}$(NUV$-r)$ CMR of late-type galaxies is different from the relatively tight $^{0.1\textrm{\protect\scriptsize K}}(u-r)$ CMR of late-type galaxies shown in Paper I.
However, $h$BLGs have a tight $^{0.1\textrm{\protect\scriptsize K}}$(NUV$-r)$ CMR, which is unusual among late-type galaxies.
This seems to be because most $h$BLGs are disc-dominated (i.e. hardly affected by bulge components; Paper I) and are not affected by an AGN.
The tight and linear $^{0.1\textrm{\protect\scriptsize K}}$(NUV$-r)$ CMR of $h$BLGs may reflect the mass -- age relation \citep{ber05,tre05}, in the sense that massive and luminous galaxies have older mean stellar ages than faint galaxies with small mass.
\citet{wyd07} showed that the UV -- optical CMR of blue sequence galaxies is not fit well using a simple linear function.
The blue sequence galaxies consist of passive blue galaxies, H{\protect\scriptsize II} blue galaxies and AGN blue galaxies in our classification scheme. Since the CMRs of H{\protect\scriptsize II} blue galaxies (i.e. $h$BEGs and $h$BLGs) are quite linear in Fig.~\ref{nuvr}, the non-linearity of the blue sequence shown by \citet{wyd07} is mainly due to the AGN blue galaxies (i.e. $s$BEGs, $l$BEGs, $s$BLGs and $l$BLGs), the CMRs of which are not linear. The fraction of the passive blue galaxies is too small to affect the CMR of the entire blue galaxies.

\subsection{Colour-colour relation}

\begin{figure*}
\includegraphics[width=168mm]{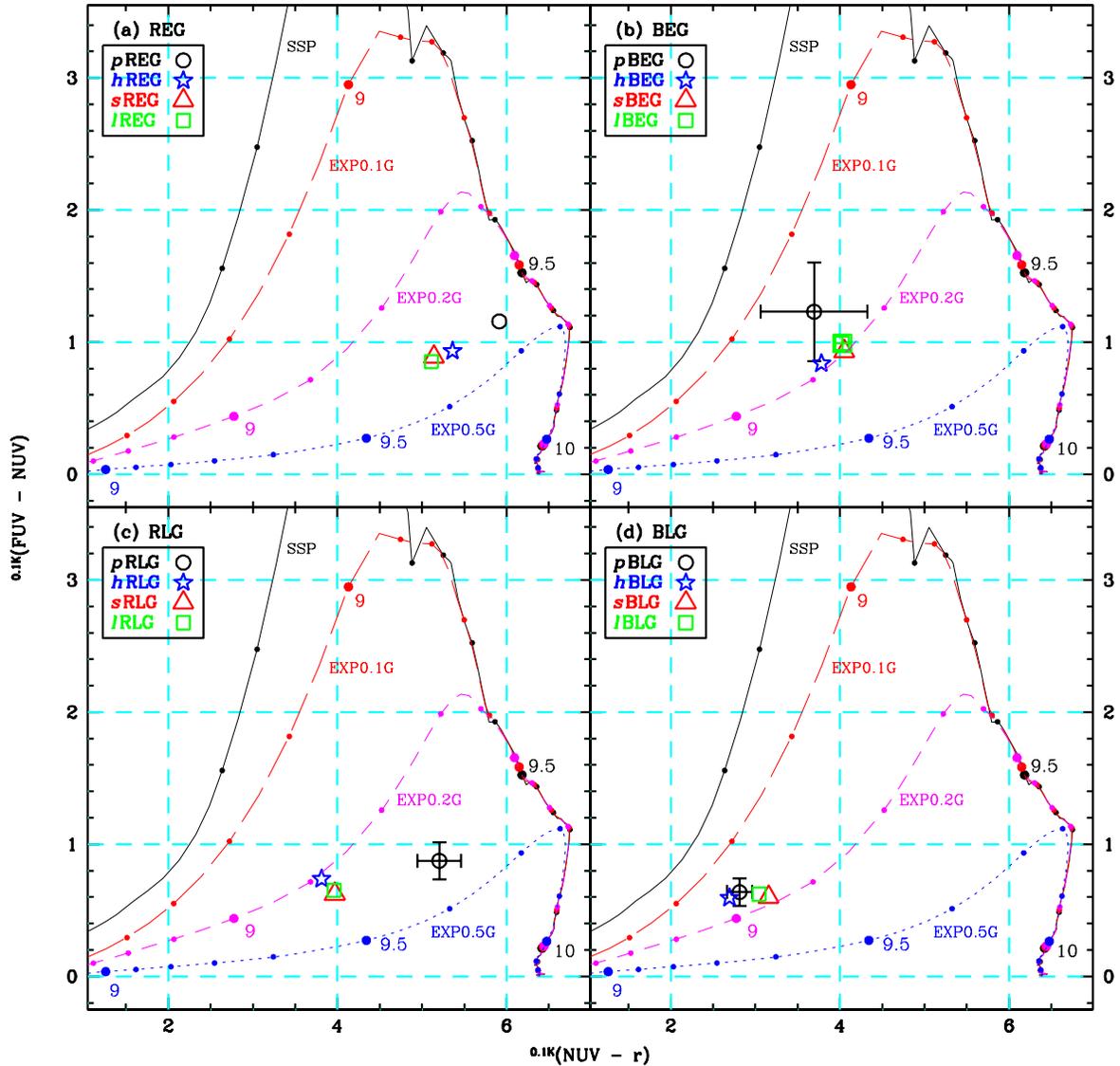}
\caption{The $^{0.1\textrm{\protect\scriptsize K}}$(FUV$-$NUV) versus $^{0.1\textrm{\protect\scriptsize K}}$(NUV$-r)$ relation of (a) REGs, (b) BEGs, (c) RLGs and (d) BLGs, in the V1 volume.
Each symbol represents the median colour of each class, and the errorbar represents the sampling error. The sampling errors for the classes without errorbars are smaller than the symbol size.
Several population synthesis models \citep{bru03} are overlaid. The solid line, long-dashed line, short-dashed line, and dotted line represent the SSP model and the exponentially-decreasing SF rate models with decreasing time-scale of 0.1, 0.2, and 0.5 Gyr, respectively. In all models, solar metal abundance (Z = 0.02) was adopted.
On the model lines, it is dotted at every 0.1 dex of age, and the numbers beside filled circles represent the log scales of the age in year since the first SF. }
\label{uvccr1}
\end{figure*}

\begin{table}
\centering
\caption{The median GALEX and SDSS colours of each class in the V1 volume\label{uvcolour}}
\begin{tabular}{rcc}
\hline \hline
& $^{0.1\textrm{\protect\scriptsize K}}$(NUV$-r)$ & $^{0.1\textrm{\protect\scriptsize K}}$(FUV$-$NUV) \\
\hline
$p$REG  &  5.83$\pm$0.06 & 1.15$\pm$0.04	\\
$h$REG  &  5.24$\pm$0.07 & 0.93$\pm$0.05	\\
$s$REG  &  4.87$\pm$0.08 & 0.85$\pm$0.05	\\
$l$REG  &  5.00$\pm$0.04 & 0.89$\pm$0.02	\\
\hline
$p$BEG  &  3.54$\pm$0.57 & 1.23$\pm$0.44	\\
$h$BEG  &  3.60$\pm$0.08 & 0.84$\pm$0.06	\\
$s$BEG  &  4.78$\pm$0.10 & 0.99$\pm$0.07	\\
$l$BEG  &  3.91$\pm$0.09 & 0.93$\pm$0.06	\\
\hline
$p$RLG  &  5.36$\pm$0.21 & 0.99$\pm$0.14	\\
$h$RLG  &  3.69$\pm$0.08 & 0.73$\pm$0.04	\\
$s$RLG  &  3.81$\pm$0.07 & 0.62$\pm$0.03	\\
$l$RLG  &  3.87$\pm$0.04 & 0.62$\pm$0.03	\\
\hline
$p$BLG  &  2.71$\pm$0.14 & 0.76$\pm$0.10	\\
$h$BLG  &  2.47$\pm$0.01 & 0.59$\pm$0.01	\\
$s$BLG  &  2.88$\pm$0.03 & 0.58$\pm$0.02	\\
$l$BLG  &  2.97$\pm$0.03 & 0.61$\pm$0.02	\\
\hline \hline
\end{tabular}
\medskip
\\The $\pm$ values are the sampling errors of the median colour.
\end{table}

Fig.~\ref{uvccr1} shows the $^{0.1\textrm{\protect\scriptsize K}}$(FUV$-$NUV) versus $^{0.1\textrm{\protect\scriptsize K}}$(NUV$-r)$ relation in each class, in the V1 volume. The median colours of each class and its sampling errors are summarised in Table \ref{uvcolour}. The axis ratio limit ($>0.6$) is applied to late-type galaxies.
Several interesting features on the UV -- optical colours of galaxies in the individual classes are found in Fig.~\ref{uvccr1}.

First, all REG classes have the median $^{0.1\textrm{\protect\scriptsize K}}$(NUV$-r)$ colour larger than 5.0, implying that the fraction of young stellar population in REGs is very small on average, compared to those of the other classes. Particularly, $p$REGs are significantly redder than non-passive REGs both in the $^{0.1\textrm{\protect\scriptsize K}}$(NUV$-r)$ colour (by more than 0.5) and in the $^{0.1\textrm{\protect\scriptsize K}}$(FUV$-$NUV) colour (by more than 0.2), showing that $p$REGs have smaller fractions of young stars than non-passive REGs. The UV -- optical colours of $s$REGs and $l$REGs are almost the same, but $h$REGs are slightly redder (by more than 0.2) than the AGN host REGs in $^{0.1\textrm{\protect\scriptsize K}}$(NUV$-r)$ colour.

Second, all BLG classes are much bluer than REG classes, both in the $^{0.1\textrm{\protect\scriptsize K}}$(NUV$-r)$ colour and in the $^{0.1\textrm{\protect\scriptsize K}}$(FUV$-$NUV) colour.
The bluest class in the UV -- optical colours is $h$BLGs.
$p$BLGs are the second-bluest class, implying that the stellar populations in $p$BLGs are very young. This confirms the result in \S\ref{S2massccr}, suggesting that $p$BLGs have been $h$BLGs up to recently or that there is current SF in the outskirts of $p$BLGs.
The AGN host BLGs are significantly redder than $h$BLGs in the $^{0.1\textrm{\protect\scriptsize K}}$(NUV$-r)$ colour (by more than 0.3), but have similar $^{0.1\textrm{\protect\scriptsize K}}$(FUV$-$NUV) colours to that of $h$BLGs within sampling errors.

Third, BEGs and non-passive RLGs have similar $^{0.1\textrm{\protect\scriptsize K}}$(NUV$-r)$ colours, which are intermediate between those of REGs and BLGs.
However, BEGs are significantly redder than non-passive RLGs in the $^{0.1\textrm{\protect\scriptsize K}}$(FUV$-$NUV) colour.
It is interesting that the same morphology classes are in the similar median $^{0.1\textrm{\protect\scriptsize K}}$(FUV$-$NUV) colour range. In other words, the median $^{0.1\textrm{\protect\scriptsize K}}$(FUV$-$NUV) colours of early-type galaxies range from 0.84 to 1.23, while those of late-type galaxies except for $p$RLGs range from 0.59 to 0.76 in Fig.~\ref{uvccr1}.
It is noted that $p$RLGs have similar colour to those of REGs, indicating that the recent SF history in $p$RLGs is similar to those in REGs. This result supports the similarity between elliptical galaxies and bulges in spiral galaxies \citep{jab07}.

In Fig.~\ref{uvccr1}, several population synthesis models \citep{bru03} are overplotted. Since the apertures of the optical and UV magnitudes are not exactly the same, there may be small offsets in the $^{0.1\textrm{\protect\scriptsize K}}$(NUV$-r)$ colours.
It is noted that all four models used in Fig.~\ref{uvccr1} have similar evolutionary path after 6 Gyr ($10^{9.8}$ yr) old, but extremely different paths before 3 Gyr ($10^{9.5}$ yr) old, showing that the UV colour is very sensitive to recent SF \citep{yi99}.
Since the $^{0.1\textrm{\protect\scriptsize K}}$(FUV$-$NUV) colour varies sensitively to the fraction of young stellar populations, it is difficult to reconstruct the multi-population SF history in each class.
For example, the colours of BEGs are explained using the EXP0.2G model, but that is not a unique solution because some models with secondary or more starbursts are available to explain the colours of BEGs.
In fact, since the mean stellar ages of BEGs estimated using the EXP0.2G model are too young compared to those estimated using optical -- NIR data (\S\ref{S2massccr}), it is plausible that BEGs have complex SF histories rather than simple EXP SF histories.
Consequently, Fig.~\ref{uvccr1} hardly gives more detailed information than that the mean stellar ages of BLGs are much younger than those of REGs and the mean stellar ages of BEGs and RLGs are intermediate.

We tested the variation of galaxy median colours in the individual classes when we use the galaxies in the V2 and in the V3 volumes instead of those in the V1 volume.
As a result, we found that the overall trends of class colour are consistent with Fig.~\ref{uvccr1}, but the sampling errors are very large due to the small size of the sample.
One notable trend is that the $^{0.1\textrm{\protect\scriptsize K}}$(FUV$-$NUV) colour range expands toward red $^{0.1\textrm{\protect\scriptsize K}}$(FUV$-$NUV), as the average optical luminosity in the volume becomes low.
For example, the $^{0.1\textrm{\protect\scriptsize K}}$(FUV$-$NUV) of $p$REGs in the V3 volume is 1.69, while that in the V1 volume is 1.15. This difference is not statistically significant due to the large sampling error in the V3 volume, but the variations of the $^{0.1\textrm{\protect\scriptsize K}}$(FUV$-$NUV) colours as the volume varies (V1 $\rightarrow$ V2 $\rightarrow$ V3) seem to be systematic.
The systematic variations of the $^{0.1\textrm{\protect\scriptsize K}}$(FUV$-$NUV) colour for REGs indicate that faint REGs are younger than bright REGs, considering the population synthesis models in Fig.~\ref{uvccr1}. This is consistent with the results using the optical and NIR data (\S\ref{S2massccr}) and the downsizing scenario \citep{cow96,ber05,tre05}.

\section{X-ray}

\begin{figure}
\includegraphics[width=84mm,height=110mm]{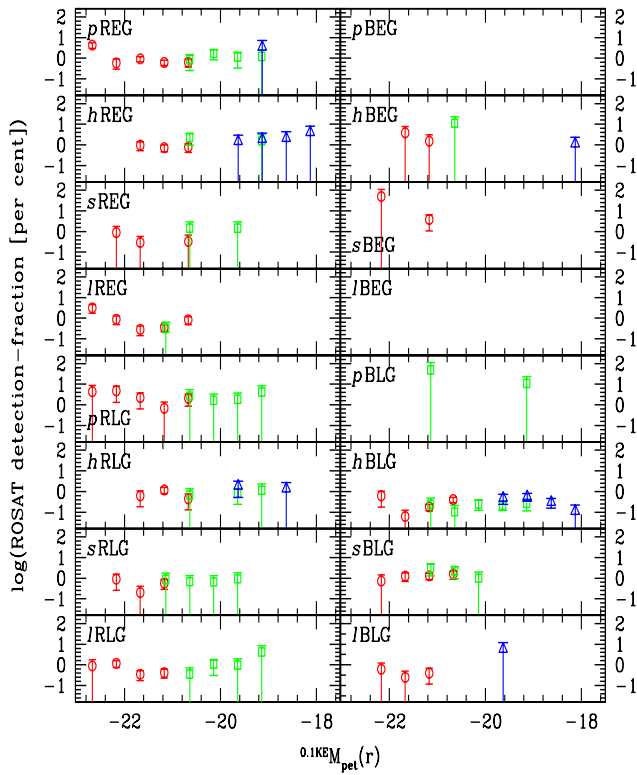}
\caption{The detection fraction of the SDSS objects in the ROSAT. The symbols are the same as those in Fig.~\ref{det2mass}.}
\label{detrosat}
\end{figure}

\begin{table}
\centering
\caption{The number of the ROSAT-detected objects in the V1 volume\label{detrosattab}}
\begin{tabular}{lrr}
\hline \hline
& REG & BEG \\
\hline
Passive &   45 (0.88 $\%$)  & 0 (0.00 $\%$)  \\
H{\protect\scriptsize II} &   18 (0.73 $\%$)  & 2 (1.35 $\%$)  \\ 
Seyfert &    3 (0.20 $\%$)  & 3 (3.06 $\%$)  \\
LINER &   27 (0.56 $\%$)  & 0 (0.00 $\%$)  \\
\hline \hline
& RLG & BLG \\
\hline
Passive &   9 (2.02 $\%$)  & 0 (0.00 $\%$)  \\
H{\protect\scriptsize II} &   13 (0.80 $\%$)  & 24 (0.25 $\%$)  \\ 
Seyfert &    7 (0.40 $\%$)  & 19 (1.29 $\%$)  \\
LINER &   17 (0.48 $\%$)  & 4 (0.31 $\%$)  \\
\hline \hline
\end{tabular}
\medskip
\\The percentages in the parentheses show the ROSAT detection fractions of the SDSS galaxies in the individual classes.
\end{table}

Table \ref{detrosattab} lists the ROSAT detection fraction of the SDSS galaxies in each class, and the dependence of the ROSAT detection fractions on optical luminosity is shown in Fig.~\ref{detrosat}.
The detection fraction of the SDSS galaxies in the ROSAT is smallest among all multi-wavelength surveys used in this paper. Most classes show detection fractions below one per cent, and the largest detection fraction is about 3 per cent (3 of 98 $s$BEGs).
It is noted that $p$RLGs show a relatively large detection fraction in the X-ray, although they are optically passive. These objects may be X-ray Bright Optically Normal Galaxies \citep[XBONGs;][]{yua04,civ07} or member galaxies in clusters of galaxies \citep{pop04}.
\citet{obr06} reported that there is a bias toward red optical colour in the ROSAT detected galaxies, which is confirmed by our result that 73 per cent of the X-ray detected objects are red galaxies, while 65 per cent of the entire galaxies in the V1 volume are red galaxies. This colour bias could be related to the environments of the X-ray luminous galaxies. The environments of those galaxies will be investigated in our next paper (in preparation).

\section{Star formation rate}\label{sfr}

\begin{table*}
\centering
\caption{Star formation rate of each fine class in the V1 volume\label{sfrtab}}
\begin{tabular}{rrrrrrr}
\hline \hline
& $\log$(SFR$_{\protect\textrm{\tiny H{\protect\tiny $\alpha$}}}$ & $\log$(SFR$_{\protect\textrm{\tiny [O{\protect\tiny II}]}}$ & $\log$(SFR$_{\protect\textrm{\tiny IRAS}}$ & $\log$(SFR$_{\protect\textrm{\tiny FIRST}}$ & $\log$(SFR$_{\protect\textrm{\tiny NVSS}}$ & $\log$(SFR$_{\protect\textrm{\tiny GALEX}}$  \\
& /M$_{\odot}$ yr$^{-1}$) & /M$_{\odot}$ yr$^{-1}$) & /M$_{\odot}$ yr$^{-1}$) & /M$_{\odot}$ yr$^{-1}$) & /M$_{\odot}$ yr$^{-1}$) & /M$_{\odot}$ yr$^{-1}$) \\
\hline
$p$REG & --- & --- & $1.78\pm0.07$ & $1.56\pm0.31$ & $1.93\pm0.40$ & $-1.36\pm0.21$ \\
$h$REG & $-1.37\pm0.40$ & $-1.33\pm0.46$ & $2.01\pm0.00$ & $1.45\pm0.24$ & $1.82\pm0.38$ & $-1.20\pm0.29$ \\
$s$REG & $-1.12\pm0.43$ & $-0.94\pm0.44$ & $1.79\pm0.12$ & $1.48\pm0.24$ & $1.75\pm0.19$ & $-1.06\pm0.25$ \\
$l$REG & $-1.29\pm0.32$ & $-1.04\pm0.38$ & $1.99\pm0.13$ & $1.53\pm0.30$ & $1.85\pm0.34$ & $-1.06\pm0.30$ \\
\hline
$p$BEG & --- & --- & --- & $1.54\pm0.00$ & $1.47\pm0.00$ & $-0.89\pm0.00$ \\
$h$BEG & $0.13\pm0.46$ & $-0.08\pm0.33$ & $1.81\pm0.39$ & $1.23\pm0.13$ & $1.70\pm0.35$ & $-0.51\pm0.23$ \\
$s$BEG & $-0.08\pm0.48$ & $-0.08\pm0.41$ & $2.03\pm0.12$ & $1.19\pm0.30$ & $1.73\pm0.16$ & $-0.72\pm0.27$ \\
$l$BEG & $-0.75\pm0.69$ & $-0.50\pm0.61$ & $2.09\pm0.17$ & $1.27\pm0.24$ & $1.90\pm0.33$ & $-0.65\pm0.22$ \\
\hline
$p$RLG & --- & --- & $1.65\pm0.01$ & $1.72\pm0.20$ & $2.49\pm0.56$ & $-1.18\pm0.45$ \\
$h$RLG & $-0.29\pm0.45$ & $-0.30\pm0.42$ & $1.79\pm0.15$ & $1.28\pm0.14$ & $1.58\pm0.17$ & $-0.55\pm0.20$ \\
$s$RLG & $-0.55\pm0.48$ & $-0.39\pm0.45$ & $1.74\pm0.17$ & $1.26\pm0.15$ & $1.59\pm0.19$ & $-0.53\pm0.22$ \\
$l$RLG & $-0.91\pm0.40$ & $-0.63\pm0.39$ & $1.88\pm0.14$ & $1.39\pm0.25$ & $1.74\pm0.31$ & $-0.50\pm0.25$ \\
\hline
$p$BLG & --- & --- & $1.57\pm0.23$ & $1.47\pm0.05$ & $1.53\pm0.04$ & $-0.08\pm0.27$ \\
$h$BLG & $0.17\pm0.27$ & $0.17\pm0.29$ & $1.80\pm0.17$ & $1.32\pm0.14$ & $1.55\pm0.15$ & $-0.12\pm0.18$ \\
$s$BLG & $0.10\pm0.40$ & $0.19\pm0.37$ & $1.91\pm0.16$ & $1.31\pm0.14$ & $1.56\pm0.13$ & $-0.19\pm0.20$ \\
$l$BLG & $-0.27\pm0.42$ & $-0.07\pm0.36$ & $1.85\pm0.14$ & $1.38\pm0.18$ & $1.62\pm0.16$ & $-0.18\pm0.20$ \\
\hline \hline
\end{tabular}
\medskip
\\Median($\log$ SFR) $\pm$ SIQR($\log$ SFR). Each SFR was estimated using each sub-sample that includes only the galaxies detected in each survey, not using the full sample. Note that the SFRs of AGN host galaxies are probably overestimated due to AGN emission.
\end{table*}

We estimate the SFRs of our sample galaxies. Since we have multi-wavelength datasets, there are several independent ways to derive SFR. \citet{hop03} presented a couple of methods for SFR estimation using the SDSS data. Among them, we adopted two spectroscopic methods using H{\protect\small $\alpha$} and [O{\protect\scriptsize II}] line luminosity, with the obscuration and aperture corrections.
In addition, we also estimated IR-SFR using the method of \citet{rie09}, radio-SFR using the method of \citet{bel03}, and UV-SFR using the method of \citet{sal07}. Dust corrections were not taken into account for the UV-SFR, because only a very small part of UV-detected objects have IR information.

\begin{table*}
\centering
\caption{Ratio of star formation rates between different surveys in the V1 volume\label{sfrrtab}}
\begin{tabular}{rrrr}
\hline \hline
& $\log$(SFR$_{\protect\textrm{\tiny IRAS}}$/SFR$_{\protect\textrm{\tiny FIRST}}$) & $\log$(SFR$_{\protect\textrm{\tiny IRAS}}$/SFR$_{\protect\textrm{\tiny NVSS}}$) & $\log$(SFR$_{\protect\textrm{\tiny IRAS}}$/SFR$_{\protect\textrm{\tiny GALEX}}$) \\
\hline
$p$REG & --- & --- & --- \\
$h$REG & --- & --- & --- \\
$s$REG & $0.53\pm0.10\;(0.14\%)$ & $0.53\pm0.00\;(0.07\%)$ & $2.77\pm0.00\;(0.07\%)$ \\
$l$REG & $0.37\pm0.20\;(0.06\%)$ & $0.57\pm0.21\;(0.10\%)$ & $3.22\pm0.00\;(0.02\%)$ \\
\hline
$p$BEG & --- & --- & --- \\
$h$BEG & $0.50\pm0.20\;(2.03\%)$ & --- & $2.97\pm0.00\;(0.68\%)$ \\
$s$BEG & $0.57\pm0.10\;(4.08\%)$ & $0.51\pm0.24\;(4.08\%)$ & $2.63\pm0.02\;(2.04\%)$ \\
$l$BEG & $0.28\pm0.00\;(0.63\%)$ & --- & $2.53\pm0.11\;(1.26\%)$ \\
\hline
$p$RLG & $0.29\pm0.00\;(0.22\%)$ & $0.09\pm0.00\;(0.22\%)$ & $2.42\pm0.00\;(0.22\%)$ \\
$h$RLG & $0.54\pm0.21\;(1.48\%)$ & $0.30\pm0.18\;(0.49\%)$ & $2.61\pm0.35\;(0.49\%)$ \\
$s$RLG & $0.30\pm0.27\;(1.98\%)$ & $0.18\pm0.19\;(0.85\%)$ & $2.32\pm0.24\;(0.56\%)$ \\
$l$RLG & $0.49\pm0.20\;(1.09\%)$ & $0.31\pm0.20\;(0.78\%)$ & $2.45\pm0.39\;(0.34\%)$ \\
\hline
$p$BLG & $-0.05\pm0.00\;(1.30\%)$ & $0.09\pm0.00\;(1.30\%)$ & $1.62\pm0.20\;(2.60\%)$ \\
$h$BLG & $0.47\pm0.17\;(1.47\%)$ & $0.26\pm0.17\;(0.92\%)$ & $1.88\pm0.28\;(1.42\%)$ \\
$s$BLG & $0.50\pm0.25\;(3.87\%)$ & $0.34\pm0.21\;(2.38\%)$ & $2.05\pm0.28\;(2.11\%)$ \\
$l$BLG & $0.41\pm0.21\;(5.15\%)$ & $0.28\pm0.16\;(3.28\%)$ & $1.84\pm0.38\;(2.34\%)$ \\
\hline \hline
\end{tabular}
\medskip
\\Median($\log$ ratio) $\pm$ SIQR($\log$ ratio). Each ratio was estimated using each sub-sample that includes only the galaxies detected both in the compared two surveys, not using the full sample. The percentage in the brackets shows the fraction of the sub-sample size to the full sample size in each fine class.
\end{table*}

Table \ref{sfrtab} summarises the SFRs derived from the SDSS, IRAS, FIRST, NVSS and GALEX data.
It should be noted that the SFRs of AGN host galaxies (i.e. Seyferts and LINERs) in most methods are inaccurate, because most SFR measures are rendered by the emission of AGN itself, such as H{\protect\small $\alpha$} line, IR luminosity, radio luminosity, and so on. It is known that [O{\protect\scriptsize II}] line is an SFR indicator that is least affected by AGN emission \citep{ho05,kim06}.

In Table \ref{sfrtab}, the SFR trends of the fine classes are different between the different surveys. For example, the SFR of REGs is much smaller than that of the other classes in the H{\protect\small $\alpha$}, [O{\protect\scriptsize II}] and GALEX results, whereas the SFR of REGs are similar to or even larger than that of the other classes in the IRAS, FIRST and NVSS results. This is because the median SFRs in each survey were estimated using each sub-sample that includes only the galaxies detected in each survey, not the full sample. That is why it is important to consider the detection fration of each class in each survey. In other words, the REGs detected in the IRAS show large SFR$_{\protect\textrm{\tiny IRAS}}$, but they are only 0.2 percent of the entire REGs.
In the H{\protect\small $\alpha$}, [O{\protect\scriptsize II}] and GALEX estimates, $h$BLGs and $h$BEGs have quite large SFR, showing their vigorous SF activity. On the other hand, the SFR of $h$REGs is very small (smaller than H{\scriptsize II} blue galaxies by $\sim1.5$ dex), indicating the SF activity in $h$REGs is very weak on average.

Table \ref{sfrrtab} summarises the SFR ratios between different surveys in log-scale: SFR$_{\protect\textrm{\tiny IRAS}}$/SFR$_{\protect\textrm{\tiny FIRST}}$,  SFR$_{\protect\textrm{\tiny IRAS}}$/SFR$_{\protect\textrm{\tiny NVSS}}$ and SFR$_{\protect\textrm{\tiny IRAS}}$/SFR$_{\protect\textrm{\tiny GALEX}}$.
Since each ratio is estimated using each sub-sample that includes only the galaxies detected both in the compared two surveys, not the full sample, it should be reminded that each value in Table \ref{sfrrtab} was derived from a very small part of the full sample. In Table \ref{sfrrtab}, it is found that the SFR$_{\protect\textrm{\tiny IRAS}}$/SFR$_{\protect\textrm{\tiny GALEX}}$ of BLGs is significantly smaller than that of RLGs, which indicates that the dust contents of RLGs are richer than those of BLGs, at least in the compared sub-samples. Passive late-type galaxies have smaller SFR$_{\protect\textrm{\tiny IRAS}}$/SFR$_{\protect\textrm{\tiny radio}}$ and SFR$_{\protect\textrm{\tiny IRAS}}$/SFR$_{\protect\textrm{\tiny GALEX}}$ than H{\scriptsize II} late-type galaxies, implying that the dust contents in passive late-type galaxies are poorer than those in H{\scriptsize II} late-type galaxies.
Since the UV, IR and radio luminosity are affected by AGN emission, it is difficult to discuss the dust contents of AGN host galaxies using the results of Table \ref{sfrrtab}.

\section{Discussion\label{DIS}}

\subsection{Red early-type galaxies}

REGs correspond to red elliptical or red S0 galaxies, which have been investigated in numerous studies \citep[e.g.][]{str01,ber05,gra09}. Typically, REGs have been known to be old and passively-evolving galaxies (which correspond to $p$REGs in our classification scheme), but Paper I showed that there are some variations of them with non-passive spectral features ($h$REGs, $s$REGs and $l$REGs). In this paper, their multi-wavelength properties show new hints on their identities.

CMR is a very useful tool to find out the stellar contents of galaxies.
The optical CMR of REGs is known to largely depend on the metallicity-magnitude relation \citep{kod97,kau98}, and Paper I showed that REGs have well-defined CMRs regardless of their spectral class.
Like the optical CMRs, the optical -- NIR CMRs of REGs in different spectral classes are quite similar, whereas the UV -- optical CMRs of REGs significantly vary depending on their spectral class.
Since it is known that optical $-$ NIR colour is sensitive to metallicity and  $\alpha$-enhancement but less sensitive to age compared to optical colour \citep{sma01,cha06}, while UV $-$ optical colour is sensitive to age \citep{yi99,dor03}, those UV--optical--NIR CMRs indicate that the different spectral classes of REGs have different mean stellar age (or even different star formation history).

Paper I showed that non-passive REGs have slightly bluer outskirts and slightly larger axis ratios than those of $p$REGs on average, which implies the existence of faint disc components in the non-passive REGs. In the UV -- optical CMR, the non-passive REGs show larger $^{0.1\textrm{\protect\scriptsize K}}$(NUV $-r)$ colour scatters (\citet{yi05} argued that such a large scatter is evidence of recent SF activity) than that of $p$REGs, whereas their optical CMR scatters do not show distinguishable difference in Paper I. From these results, we infer that the SF disc components in the non-passive REGs may be so faint that they hardly affect their total optical colours, but they affect their total UV -- optical colours. The SFRs of non-passive REGs are very small compared to the other non-passive galaxies in every SFR estimation.

The colour--colour diagrams provide more information about the stellar populations in each spectral class of REGs.
In the $^{0.1\textrm{\protect\scriptsize K}}(u-r)$ -- $^{0.1\textrm{\protect\scriptsize K}}(r-K_{s})$ diagram, the four spectral classes of REGs have similar colours, but $p$REGs have slightly bluer $^{0.1\textrm{\protect\scriptsize K}}(r-K_{s})$ colour (by $0.03-0.04$) than those of non-passive REGs. This indicates that $p$REGs are slightly metal-poorer than non-passive REGs, which is also verified using the metallicity estimates of \citet{gal06}: an offset of  $\Delta$log Z $\sim0.02$ is found between $p$REGs and $h$REGs on average.
On the other hand, $p$REGs are significantly redder than non-passive REGs in the $^{0.1\textrm{\protect\scriptsize K}}$(FUV$-$NUV) and $^{0.1\textrm{\protect\scriptsize K}}$(NUV$-r)$ colours, suggesting that non-passive REGs have more young stars than $p$REGs. The relatively large metallicity and young age of the non-passive REGs indicate recent star formation events that produced young and metal-rich stellar populations in the non-passive REGs.

Previous studies showed that several dust features (e.g. patches, lanes and filaments) are often found in nearby elliptical galaxies \citep[e.g.][]{van95,lee04,tem07}.
However, REGs are hardly detected in the MIR and FIR bands, showing that REGs have little dust. At first glance, such silence of REGs in the IR bands seems not to agree with our knowledge about nearby elliptical galaxies, but this may be due to the poor detection efficiency of IRAS. In other words, the IR emission from the dust features in REGs may not be sufficiently strong to be detected at $0.02<z<0.10$ by IRAS.

REGs show conspicuous dependence of their radio luminosity on their optical absolute magnitude, in the sense that bright REGs are detected in the radio more efficiently than faint REGs.
Since \citet{bes05b} and \citet{cro07} showed that the fraction of radio-loud AGN host galaxies is a strong function of stellar mass (i.e. radio-loud AGN host galaxies tend to be massive galaxies), this trend indicates that the major radio source in REGs is radio-loud AGNs. This idea is also supported indirectly by the silence of REGs in the IR bands, because, if the radio emission originates from SF regions, not from AGNs, they will be accompanied by a considerable amount of dust that is easily detected in the IR bands.
Some radio-loud REGs are optically passive, corresponding to previously known \emph{radio-loud AGNs without emission line} \citep[e.g.][]{kau08}.

In summary, (a) non-passive REGs have younger mean stellar age and larger mean metallicity than $p$REGs, and (b) the SFRs of non-passive REGs are so small in every SFR estimation that they hardly affect their total optical colours, but they affect their total UV -- optical colours. (c) These results indicate that recent star formation events have produced young and metal-rich stellar populations in the non-passive REGs. (d) The major radio source in REGs seems to be radio-loud AGNs, and some radio-loud REGs are optically passive.

\subsection{Blue early-type galaxies}

BEGs have bluer colour (particularly, bluer centre) than REGs \citep{men01,lee06,cho07}. Their morphology is similar to REGs, but it is known that BEGs tend to be fainter than REGs on average at low and intermediate redshift \citep{im01}. In Paper I, the structural features of $h$BEGs, $p$BEGs and $p$REGs are similar, possibly forming the evolutionary sequence of $h$BEGs $\rightarrow$ $p$BEGs $\rightarrow$ $p$REGs, as suggested in \citet{lee06,lee07}. Multi-wavelength properties in this paper show more detailed clues to their stellar contents.

The most outstanding feature in the optical -- NIR CMR of BEGs is the rapid slope of the $h$BEG CMR. While the bright $h$BEGs are similar to the bright REGs in their $^{0.1\textrm{\protect\scriptsize K}}(r-K_{s})$ colour, the faint $h$BEGs are bluer than faint REGs, in Fig.~\ref{Lrk}.
According to the age-metallicity estimates of \citet{gal06}, both age and metallicity affect the rapid CMR slope of $h$BEGs, but the effect of metallicity seems to be larger than that of age.
In other words, faint BEGs tend to be significantly metal-poorer than faint REGs, whereas bright BEGs is similar to bright REGs in their mean metallicity. This result indicates that bright BEGs have evolved more rapidly (i.e. faster metal enrichment) than faint BEGs.

The UV -- optical -- NIR colours provide hints on the star formation history of BEGs. BEGs seem to be much younger than REGs, when comparing their colours to the population synthesis models. In the optical -- NIR colours, $p$BEGs and $l$BEGs are close to the SSP model, while $h$BEGs and $s$BEGs are well described by the SSP + recent SF model.
The IRAS detection fractions of $p$BEGs and $l$BEGs are small, compared to those of BLGs. On the other hand, the IRAS detection fractions and estimated SFRs of $h$BEGs and $s$BEGs are almost as large as those of BLGs. This indicates that $h$BEGs and $s$BEGs have very young or currently-forming stars, probably accompanied by rich dust reservoirs.

In summary, (a) faint BEGs tend to be significantly metal-poorer than faint REGs, whereas bright BEGs is similar to bright REGs in their mean metallicity, showing that bright BEGs have evolved more rapidly than faint BEGs. (b) The UV -- optical -- NIR colours of BEGs are well explained using the SSP + recent SF model, without significant difference in their formation epoch between different spectral classes. (c) The IRAS detection fraction and the estimated SFRs show that $h$BEGs and $s$BEGs have very young or currently-forming stars, probably accompanied by rich dust reservoirs.

\subsection{Red late-type galaxies}

The existence of RLGs have been noted in several previous studies \citep[e.g.][]{cho07}, but intensive studies on their properties are rare.
Paper I showed that about $37\%$ of entire late-type galaxies are RLGs in the V1 volume (this fraction decreases to $12\%$ in the V3 volume).
In Paper I, RLGs seem to consist of two kinds of late-type galaxies: bulge-dominated late-type galaxies and edge-on late-type galaxies. That is, both large bulge fraction and internal extinction due to high disc inclination make late-type galaxies red.
The results using the IRAS data are useful to understand the dust contents of RLGs. In the MIR and FIR bands, RLGs are detected more efficiently compared to early-type galaxies, but the IRAS detection fraction of RLGs is smaller than that of BLGs, showing that \emph{dust extinction may not be a dominant factor making RLGs red}.

It is noted that the estimated SFR$_{\textrm{\tiny IRAS}}$/SFR$_{\textrm{\tiny GALEX}}$ of $h$RLGs ($2.61\pm0.35$) is significantly larger than that of $h$BLGs ($1.88\pm0.28$), which seems to be evidence for RLGs dust-richer than BLGs. However, those values were estimated using very small part of the full sample: 8 $h$RLGs ($0.49\%$) and 137 $h$BLGs ($1.42\%$). Reminding that the IRAS detection fraction is much smaller than the GALEX detection fraction, the objects detected both in the IRAS and GALEX may be biased to the dust-richest galaxies in the sample. In that case, the comparison between the $0.49\%$ of $h$RLGs and the $1.42\%$ of $h$BLGs is unfair. Actually, the median SFR$_{\textrm{\tiny IRAS}}$/SFR$_{\textrm{\tiny GALEX}}$ of the dust-richest $0.49\%$ of $h$BLGs is $2.36\pm0.15$, which is not significantly smaller than the value of $h$RLGs.

The detection fractions of RLGs in the UV -- NIR bands reflect their stellar contents. The NIR detection fraction of RLGs is as large as that of REGs, but the UV detection fraction of RLGs is intermediate between REGs and blue galaxies. This indicates that \emph{the young star fraction of RLGs is larger than that of REGs but smaller than those of BEGs or BLGs}. The SFR of $h$RLGs estimated in any method is smaller than that of $h$BLGs, reflecting their relatively small SF disc.

More detailed clues to the stellar contents of RLGs are found in the multi-wavelength colours.
In the $^{0.1\textrm{\protect\scriptsize K}}(u-r)$ -- $^{0.1\textrm{\protect\scriptsize K}}(r-K_{s})$ colour-colour diagram (Fig.~\ref{optnir1}), RLGs seem to have a considerable amount of old stellar populations as well as young stellar populations, which is consistent with the idea of Paper I that RLGs are bulge-dominated late-type galaxies.
The young stars in RLGs are more outstanding in the UV bands. Non-passive RLGs are much bluer than REGs in the $^{0.1\textrm{\protect\scriptsize K}}$(FUV$-$NUV) and $^{0.1\textrm{\protect\scriptsize K}}$(NUV$-r)$ colours, in spite that REGs and RLGs were selected using the same optical colour criterion.
Since the UV -- optical colour is sensitive to young stars, this indicates that non-passive RLGs have more young stars than REGs.

In Paper I, $p$RLGs show optical properties similar to REGs rather than non-passive RLGs. Such a trend is also found in their multi-wavelength properties. For example, $p$RLGs have similar UV -- optical -- NIR colours to those of REGs, suggesting that the stellar contents of $p$RLGs are comparable to those of REGs. Moreover, the detection fraction of $p$RLGs in the radio shows a similar trend to those of REGs rather than non-passive RLGs. That is, optically bright $p$RLGs are detected in the radio with significantly higher efficiency than optically faint $p$RLGs, which indicates that the major radio source in $p$RLGs is radio-loud AGNs. On the other hand, the radio detection fraction of non-passive RLGs does not strongly depend on their absolute magnitude, showing that the major radio source in non-passive RLGs may be SF.
These results support the idea of Paper I that $p$RLGs are the intermediate objects between early-type galaxies and late-type galaxies: extremely bulge-dominated late-type galaxies.

In summary, (a) there is no clear evidence that RLGs are dust-richer than BLGs, and (b) they seem to have intermediate star formation between REGs and blue galaxies. (c) The UV -- optical -- NIR colours indicate that RLGs have not only a considerable amount of old stars but also more young stars than REGs. (d) As they do in the optical properties, $p$RLGs show properties similar to REGs in their multi-wavelength colours and radio detection trend.

\subsection{Blue late-type galaxies}

BLGs correspond to well-known spiral or irregular galaxies typically with blue colour, low light-concentration and vigorous SF. Paper I showed that most BLGs are star-forming ($h$BLGs), but a very small fraction of BLGs have no evidence of current star formation ($p$BLGs). It was also shown that there are some BLGs with AGN ($s$BLGs, $l$BLGs). In this paper, some previously known properties are confirmed using multi-wavelength data, and new hints on the identity of $p$BLGs are found.

BLGs are relatively poorly-detected in the NIR bands, while their detection fraction in the UV, MIR, FIR and radio is largest among all classes. This reflects the fact that BLGs are mostly late-type galaxies with vigorous SF. 
The SFR$_{\protect\textrm{\tiny H{\protect\tiny $\alpha$}}}$ and SFR$_{\protect\textrm{\tiny [O{\protect\tiny II}]}}$ of $h$BLGs are largest among all classes. SFR$_{\protect\textrm{\tiny H{\protect\tiny $\alpha$}}}$($h$BLG) is larger than SFR$_{\protect\textrm{\tiny H{\protect\tiny $\alpha$}}}$($h$REG) by 1.7 dex; larger than SFR$_{\protect\textrm{\tiny H{\protect\tiny $\alpha$}}}$($h$RLG) by 0.7 dex; and larger than SFR$_{\protect\textrm{\tiny H{\protect\tiny $\alpha$}}}$($h$BEG) by 0.2 dex. However, in the estimation using the IR, radio and UV, the SFR of BLGs is not particularly larger than the other classes' SFRs, due to the selection effect (described in \S\ref{sfr}).

The CMRs in different wavelength datasets reveal the dependence of BLGs' stellar population on their luminosity.
While the optical CMR slope of BLGs is significantly larger than that of REGs (Paper I), the $^{0.1\textrm{\protect\scriptsize K}}(r-K_{s})$ CMR slope of BLGs is similar to those of REGs in Fig.~\ref{Lrk}. Since optical -- NIR colour is known to be less sensitive to the stellar age than optical colour \citep{sma01}, this variation of the CMR slope according to wavelength indicates that the age difference between bright BLGs and faint BLGs is significantly responsible for the rapid CMR slope of BLGs in the optical band.
Furthermore, $h$BLGs show a tight and linear UV CMR, which is even tighter than that of REGs. Since UV light is very sensitive to stellar age, the tight UV CMRs of $h$BLGs reflects their well-defined luminosity -- age relation.
These results show that faint BLGs are younger than bright BLGs. From this, the previously known mass -- age relation \citep[or galaxy downsizing;][]{cow96,ber05,tre05} is confirmed using multi-wavelength data.

$p$BLGs are optically passive, but their UV -- optical -- NIR colours indicate their mean stellar ages are quite young. Thus, if $p$BLGs are intrinsically passive, their SF activity may have stopped recently. Otherwise, there may be current SF in the outskirts of $p$BLGs that makes $p$BLGs blue in the optical and UV bands, but their emission lines are not detected in the SDSS fibre spectroscopy. The detection fraction of $p$BLGs (3.90$\%$) larger than that of $h$BLGs (2.77$\%$) shows a possibility that $p$BLGs are dusty obscured $h$BLGs, but the uncertainty is too large.

In summary, (a) BLGs are mostly late-type galaxies with vigorous SF and their SFR is largest among all classes. (b) The previously known mass -- age relation of galaxies (i.e. galaxy downsizing) is confirmed in BLGs using multi-wavelength data, in the sense that faint BLGs are younger than bright BLGs. (c) If $p$BLGs are intrinsically passive, their SF activity may have stopped very recently. Otherwise, there may be current SF in the outskirts of $p$BLGs, out of the range of the SDSS spectroscopy fibre.

\subsection{Comparison in a given spectral class}

Passive galaxies consist of four fine classes in this paper: $p$REGs, $p$BEGs, $p$RLGs and $p$BLGs. $p$REGs show the properties of typical red elliptical galaxies that have evolved passively. On the other hand, $p$BEGs have blue colours and probably have suffered recent SF, although they do not show line emissions currently. Meanwhile, $p$RLGs have properties similar to those of $p$REGs and there is no evidence that they have suffered recent SF, but $p$RLGs are slightly more diffuse than $p$REGs (Paper I). $p$BLGs seem to be late-type galaxies that have suffered very recent SF quenching, or to be late-type galaxies with current SF in their outskirts.

The four H{\protect\scriptsize II} fine classes -- $h$REGs, $h$BEGs, $h$RLGs and $h$BLGs -- commonly show evidence for current SF. However, Paper I presented that the spatial distribution of SF is different among those fine classes; that is, the SF of $h$BEGs is relatively concentrated on their centres, while the SF in the other H{\protect\scriptsize II} galaxies is mainly distributed in their outskirts.
The three H{\protect\scriptsize II} fine classes with outskirt-biased SF -- $h$REGs, $h$RLGs and $h$BLGs -- show also different properties from each other in this paper.
For example, the SF features of $h$REGs are hardly found in the relatively shallow surveys: IRAS, FIRST and NVSS, which implies that the SF activities in $h$REGs are very weak. In addition, The UV -- optical CMR of $h$RLGs seems not to be linear, unlike that of $h$BLGs. Different bulge-to-disc ratio among the fine classes may be responsible for those differences.

In the comparison of SFR$_{\protect\textrm{\tiny [O{\protect\tiny II}]}}$ that is least affected by AGN emission \citep{ho05,kim06} between AGN host galaxies (i.e. Seyfert galaxies and LINER galaxies), Seyfert galaxies always have larger median SFR$_{\protect\textrm{\tiny [O{\protect\tiny II}]}}$ than LINER galaxies in a given morphology-colour class. Since LINERs are less active AGNs than Seyferts, this result means that more active AGNs (Seyferts) have more active SF in their host galaxies.
However, according to the AGN feedback scenario, AGN activity tends to suppress SF activity \citep{ant08,raf08}, which seems to contradict to our results if Seyferts and LINERs are independent AGN types. Thus, if we adopt the AGN feedback scenario, Seyferts and LINERs may be evolutionarily connected; for example, LINER galaxies may be at the end of the AGN feedback process, the SF of which may have been already sufficiently suppressed by AGN activity at the Seyfert phase. Evidence of such evolutionary connections between different types of AGNs has been found in several recent studies \citep[e.g.][]{sch07,sch09,hic09}.

Some AGN host galaxy classes sometimes show different properties even though they are of the same AGN type. For example, the differences in the median IRAS colours between $s$BEGs and $s$RLGs, and between $l$REGs and $l$BLGs are significantly large ($>3\sigma$), which shows their different SEDs. The UV -- optical CMRs are also obviously different between the fine classes of AGN host galaxies: AGN host early-type galaxies except for $l$BEGs have negative CMR slopes, while $l$BEGs and AGN host late-type galaxies have positive CMR slopes.
We can not fully explain what results in those differences among the fine classes in this paper. However, it is clear that such details could not even be found if galaxies were classified using simple schemes.

\section{Conclusions}

The multi-wavelength properties of galaxies in the fine classes provide evidence supporting the main results in Paper I, and add several new findings on them. 
Since most main results are digested in the Discussion section (\S\ref{DIS}), we simply list the important findings in this paper here.

\begin{itemize}
\item[(1)] From the UV -- optical -- NIR colours, non-passive REGs seem to have larger metallicity and younger age than $p$REGs. This implies that non-passive REGs may have suffered recent SF events, which have produced young and metal-rich stellar populations in them.
\end{itemize}

\begin{itemize}
\item[(2)] REGs show conspicuous dependence of their radio luminosity on their optical absolute magnitude, which indicates that the major radio source in REGs is radio-loud AGNs. Some radio-loud REGs are optically passive, corresponding to previously known \emph{radio-loud AGNs without emission line}.
\end{itemize}

\begin{itemize}
\item[(3)] $h$BEGs have the most negative slopes among all classes in the optical -- NIR CMR. Both age and metallicity may affect the rapid CMR slope of $h$BEGs, but the effect of metallicity seems to be larger than that of age.
\end{itemize}

\begin{itemize}
\item[(4)] The UV -- optical -- NIR colours of BEGs are well explained using the SSP + recent SF model, if it is supposed that there is no significant difference in their formation epoch between different spectral classes.
\end{itemize}

\begin{itemize}
\item[(5)] The IRAS detection fractions of $p$BEGs and $l$BEGs are small, compared to those of BLGs, but the IRAS detection fractions of $h$BEGs and $s$BEGs are as large as those of BLGs. This shows that $h$BEGs and $s$BEGs have very young or currently-forming stars, probably accompanied by rich dust reservoirs.
\end{itemize}

\begin{itemize}
\item[(6)] RLGs have intermediate star formation between REGs and blue galaxies. The UV -- optical -- NIR colours indicate that RLGs have not only a considerable amount of old stars but also more young stars than REGs.
\end{itemize}

\begin{itemize}
\item[(7)] The UV -- optical colours and the radio detection trend of $p$RLGs show that $p$RLGs have properties similar to REGs rather than non-passive RLGs.
\end{itemize}

\begin{itemize}
\item[(8)] The IRAS detection fraction of RLGs is smaller than that of BLGs, implying that dust extinction may not be the dominant factor making RLGs red.
\end{itemize}

\begin{itemize}
\item[(9)] Not only the CMR slope variation between the optical and NIR bands but also the tight and linear UV CMR of $h$BLGs shows that faint BLGs are younger than bright BLGs. From this, the previously known mass -- age relation (or galaxy downsizing) is confirmed using multi-wavelength data.
\end{itemize}

\begin{itemize}
\item[(10)] The UV -- optical -- NIR colours indicate that the mean stellar ages of $p$BLGs are quite young. Therefore, if $p$BLGs are intrinsically passive, their SF activity may have stopped recently. Otherwise, there may be current SF in the outskirts of $p$BLGs.
\end{itemize}

\begin{itemize}
\item[(11)] The SFR$_{\protect\textrm{\tiny [O{\protect\tiny II}]}}$ of Seyferts is always larger than that of LINERs in a given morphology-colour class. Considering the AGN feedback, it is a possible scenario that LINER galaxies may be at the end of the AGN feedback process, the SF of which may have been already sufficiently suppressed by AGN activity at the Seyfert phase.

\end{itemize}

\begin{itemize}
\item[(12)] Among AGN host galaxies, some fine classes sometimes show different properties, such as their IRAS colours or UV -- optical CMRs, even though they belong to the same spectral class. The physical origin of such differences is an open question.
\end{itemize}

This paper is the second in the series of comprehensive studies on the nature of the SDSS galaxies in finely-divided classes.
In the following paper \citep{lee09}, we will inspect the environments of the SDSS galaxies divided into the fine classes.

\section*{Acknowledgements}

The authors thank the anonymous referee for very useful comments that
improved significantly the original manuscript.
JHL appreciates the support and advice of Dr. Eon-Chang Sung.
This work was supported in part by a grant (R01-2007-000-20336-0) from the Basic Research Program of the Korea Science and Engineering Foundation (KOSEF).
CBP and YYC acknowledge the support of the KOSEF through the Astrophysical Research Centre for the Structure and Evolution of the Cosmos (ARCSEC).

Funding for the SDSS and SDSS-II has been provided by the Alfred P. Sloan Foundation, 
the Participating Institutions, the National Science Foundation, 
the US Department of Energy, the National Aeronautics and Space Administration, 
the Japanese Monbukagakusho, the Max Planck Society, and the Higher Education Funding Council for England.
The SDSS Web site is http://www.sdss.org/.
The SDSS is managed by the Astrophysical Research Consortium for the Participating Institutions. 
The Participating Institutions are the American Museum of Natural History, 
Astrophysical Institute Potsdam, the University of Basel, the University of Cambridge, 
Case Western Reserve University, the University of Chicago, Drexel University, Fermilab, 
the Institute for Advanced Study, the Japan Participation Group, Johns Hopkins University, 
the Joint Institute for Nuclear Astrophysics, the Kavli Institute for Particle Astrophysics and Cosmology, 
the Korean Scientist Group, the Chinese Acadeour of Sciences (LAMOST), Los Alamos National Laboratory, 
the Max-Planck-Institute for Astronomy (MPIA), the Max Planck Institute for Astrophysics (MPA), 
New Mexico State University, Ohio State University, the University of Pittsburgh, the University of Portsmouth, 
Princeton University, the US Naval Observatory, and the University of Washington.

The FIRST Survey is supported in part under the auspices of the Department of Energy by Lawrence Livermore National Laboratory under contract no. W-7405-ENG-48 and the Institute for Geophysics and Planetary Physics.
This publication makes use of data products from the 2MASS, which is a joint project of the University of Massachusetts and the Infrared Processing and Analysis Centre/California Institute of Technology, funded by the National Aeronautics and Space Administration and the National Science Foundation.
The galaxy Evolution Explorer (GALEX) is a NASA Small Explorer. The mission was developed in cooperation with the Centre National d'Etudes Spatiales of France and the Korean Ministry of Science and Technology.

\label{lastpage}

\end{document}